\documentclass[aps,rmp,reprint,amsmath,amssymb,graphicx,longbibliography,superscriptaddress,floatfix]{revtex4-1}
\usepackage{graphicx}
\usepackage{xcolor}
\usepackage{ulem}
\usepackage{bm}
\usepackage{slashed}

\usepackage{ marvosym }  

\usepackage{hyperref}
\hypersetup{colorlinks,%
citecolor=black,%
filecolor=black,%
linkcolor=black,%
urlcolor=blue}

\newcommand{\ud}{\mathrm{d}}

\newcommand{\be}{\begin{equation}}
\newcommand{\ee}{\end{equation}}
\newcommand{\ba}{\begin{eqnarray}}
\newcommand{\ea}{\end{eqnarray}}
\newcommand{\la}{\langle}
\newcommand{\ra}{\rangle}

\newcommand{\red}[1]{{\color{red} #1}}

\newcommand{\uncomment}[1]{{\red{[ .... ]}}}

\allowdisplaybreaks
\begin{document}

\title{Colloquium: Gravitational Form Factors of the Proton}

\author{V.~D.~Burkert}
\affiliation{Thomas Jefferson National Accelerator Facility,\unpenalty~Newport News,\unpenalty~VA,\unpenalty~USA}

\author{L.~Elouadrhiri}
\affiliation{Thomas Jefferson National Accelerator Facility,\unpenalty~Newport News,\unpenalty~VA,\unpenalty~USA}
\affiliation{Center of Nuclear Femtography,\unpenalty~Newport News,\unpenalty~VA,\unpenalty~USA}

\author{F.~X.~Girod}
\affiliation{Thomas Jefferson National Accelerator Facility,\unpenalty~Newport News,\unpenalty~VA,\unpenalty~USA}

\author{C.~Lorc\'e}
\affiliation{CPHT,\unpenalty~CNRS,\unpenalty~\'Ecole polytechnique,\unpenalty~Institut Polytechnique de Paris,\unpenalty~91120 Palaiseau,\unpenalty~France}

\author{P.~Schweitzer}
\affiliation{Department of Physics,\unpenalty~University of Connecticut,\unpenalty~Storrs,\unpenalty~CT 06269,\unpenalty~USA}

\author{P.~E.~Shanahan}
\affiliation{Center for Theoretical Physics,\unpenalty~Massachusetts Institute of Technology,\unpenalty~Cambridge,\unpenalty~MA 02139,\unpenalty~USA}

\date{\today}  
\begin{abstract}
The physics of the gravitational form factors
of the proton, and their understanding within
quantum chromodynamics, has
advanced significantly in the past two decades through both theory and experiment.
This Colloquium provides an overview of this progress,
highlights the physical insights unveiled by studies
of gravitational form factors, and reviews their interpretation in terms of the mechanical properties of the proton.
\end{abstract}


\maketitle
\tableofcontents{}

\section{Introduction}
\label{Sec-1:intro} 

This Colloquium reviews the recent theoretical 
and experimental progress in studies of the 
gravitational form factors of the proton and other 
hadrons, which has shed fascinating new light on 
the proton's structure and its mechanical properties. 
To place this emerging area in context, the history 
of proton structure and its description in 
quantum chromodynamics are first reviewed.

\vspace{-4mm}

\subsection{Anomalous magnetic moment} \vspace{-2mm}
\label{Sec-1A:magnetic-moment}
 
Soon after the proton \cite{Rutherford:1919fnt} and neutron 
\cite{Chadwick:1932ma} were established as the constituents 
of atomic nuclei, experiments showed that these spin-$\frac12$
particles with nearly equal masses 
$M_N \simeq 940\,{\rm MeV}/c^2$ are not pointlike 
elementary fermions. If they were, the Dirac equation would 
predict the magnetic moment of the proton to be one nuclear 
magneton  $\mu_N\equiv e\hbar/(2M_N)$ and that of an 
electrically neutral particle like the neutron to be zero. 
Instead, the proton magnetic moment was measured to be about 
$\mu_p \simeq 2.5\,\mu_N$ \cite{Frisch-Stern}. Later, the 
neutron magnetic moment was found to be 
$\mu_n \simeq -1.5\,\mu_N$ \cite{Alvarez:1940zz};
for the modern values of the magnetic moments see \cite{ParticleDataGroup:2022pth}.
These experiments have shown that the 
nucleon is not a pointlike elementary particle,  
giving birth in 1933 to the field of proton structure. 

Protons and neutrons are hadrons, particles that feel the 
strong force, which is the strongest interaction known in 
nature. Based on approximate isospin symmetry, they are 
understood as partnered (isospin up/down) states, referred 
to collectively as the nucleon \cite{Heisenberg:1932dw}.
As the constituents of nuclei, nucleons are responsible 
for more than $99.9\,\%$ of the mass of matter in the 
visible universe, and have naturally become the most 
experimentally studied objects in hadronic physics.

\begin{figure}
\includegraphics[width=0.9\columnwidth]{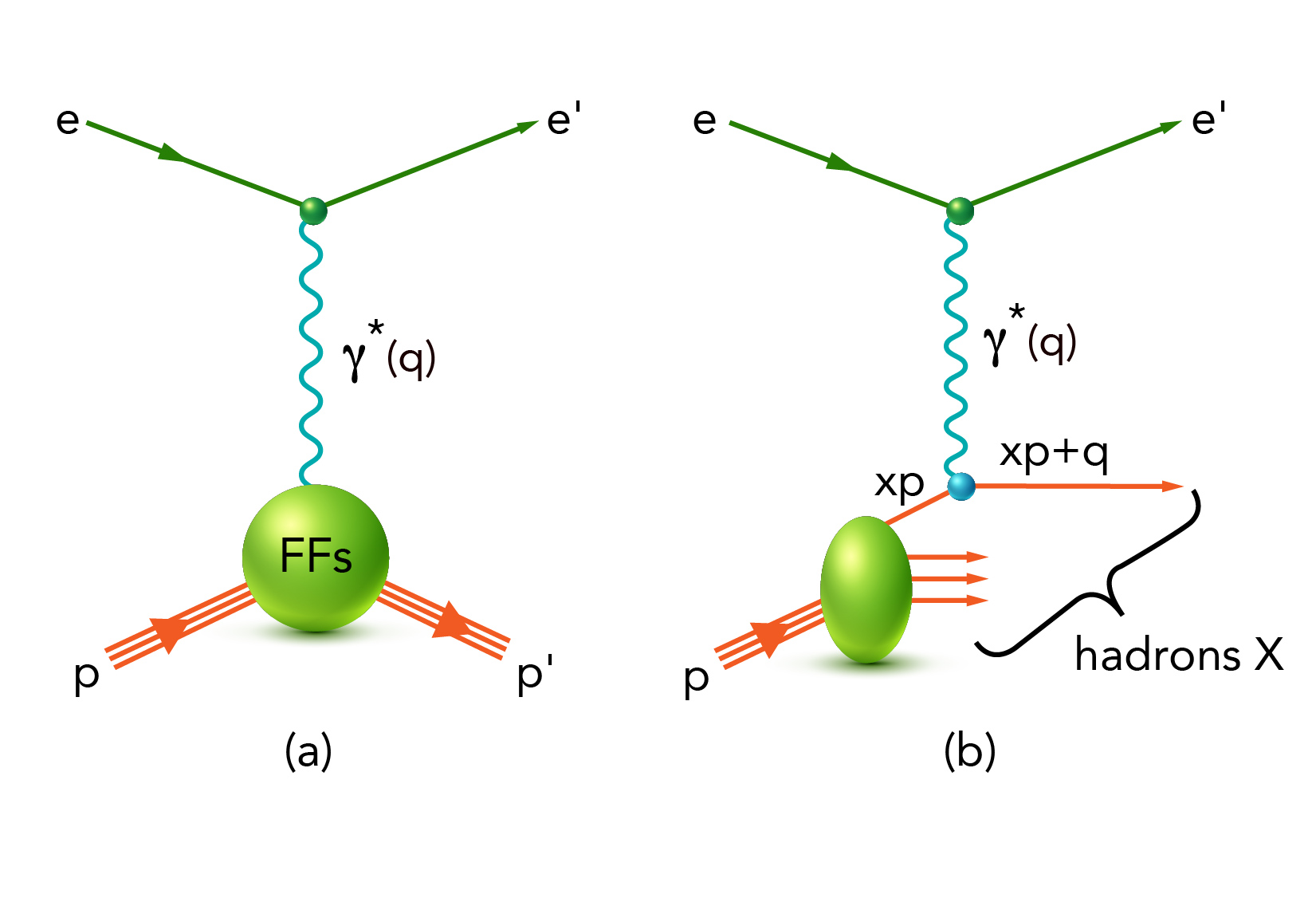}

\vspace{-6mm}

\caption{\label{Fig1}
(a) The elastic electron-proton scattering process
in which the electromagnetic form factors (FFs)
are measured.
(b) Inclusive deep inelastic scattering (DIS) 
where the proton is dissociated into a 
final state consisting of unresolved hadrons. 
In the Bjorken limit $p\cdot q\to \infty$ 
and $Q^2 = -q^2 \to \infty$ with 
$x_B=Q^2/(2p\cdot q)$ fixed, DIS 
is interpreted  
in the so-called infinite-momentum frame 
as the scattering of electrons off pointlike
quarks carrying the fraction $x$ of the nucleon's
momentum, where $x=x_B$ up to corrections 
suppressed by $M_N^2/Q^2$.}
\end{figure}

\vspace{-4mm}

\subsection{The proton's finite size}\vspace{-2mm}
\label{sect-I.B.}

An important milestone in the field of nucleon structure 
was brought by studies of elastic electron-proton 
scattering, shown in Fig.~\ref{Fig1}a,
which revealed early insights into the proton's size.
The deviations in scattering data from expectations for pointlike particles are encoded in terms of  form factors (FFs) defined 
through matrix elements of the electromagnetic current operator,
$\la p',\vec{s}^{\,\prime} |J^\mu_{em}|p,\vec{s}\ra$,
where $|p,\vec{s}\ra$ is the initial state of the proton with momentum $p$ polarized along the $\vec{s}$ direction, and analogously for the final proton state. 

These FFs would be constants for pointlike particles,
but they were found to be pronounced functions of the 
Mandelstam variable $t=(p^\prime-p)^2$. A spin-$\frac12$ 
particle has two electromagnetic FFs, $F_1(t)$ and $F_2(t)$,
defined such that $F_1(0)$ is the electric charge in units of 
$e$, and $F_2(0)$ is the anomalous magnetic moment, i.e.,\ the
deviation from the value predicted by the Dirac equation, in 
units of $\mu_N$. 
Knowledge of the $t$-dependence of electromagnetic FFs
allowed information about the spatial distributions
of electric charge and magnetization to be inferred 
\cite{Sachs:1962zzc}
(more discussion of this interpretation can be found in~\cite{Lorce:2020onh,Chen:2022smg,Chen:2023dxp}). 
This led to the first determination of the proton charge 
radius of $(0.74\pm 0.24)\,$fm \cite{Mcallister:1956ng}. 
These experiments have continued to this day, and, using 
a variety of experimental techniques, they resulted in 
a much more precise knowledge of the proton's charge 
radius~\cite{ParticleDataGroup:2022pth}.

\subsection{Discovery of partons}
\label{Sect_IC}
The 1950s witnessed immense progress in accelerator and detection
techniques followed by a proliferation of discoveries of strongly 
interacting particles and resonances, including particles like the 
antiproton, 
$\Delta$, 
and $\Xi$, 
see the early review \cite{EarlyReview1961}.
On the theory side, this led to the development of the quark 
model \cite{Gell-Mann:1964ewy,Zweig:1964ruk} in which hadrons 
are classified according to quantum numbers which are understood 
to arise from various combinations of ``quarks''. The ``quarks'' 
in this model were group-theoretical objects, and their dynamics 
were unknown.

The next milestone was brought by high-energy experiments carried
out at the Stanford Linear Accelerator, where the Bjorken scaling 
predicted on the basis of current algebra and dispersion relation 
techniques \cite{Bjorken:1968dy} was observed in inclusive deep 
inelastic scattering (DIS) \cite{Bloom:1969kc}.
The response of the nucleon in DIS is described by structure
functions which, on general grounds, are functions of the 
Lorentz invariants $p\cdot q$ and $Q^2 = -q^2$, where $p^\mu$ is 
the nucleon four-momentum and $q^\mu$ the four-momentum transfer, 
see Fig.~\ref{Fig1}b. Bjorken scaling is the property that, 
in the high-energy limit $p\cdot q\to\infty$ and $Q^2\to\infty$
with their ratio fixed,
the structure functions are, to a first approximation, functions
of a single variable $x_B=Q^2/(2p\cdot q)$ which on kinematical 
grounds satisfies $0<x_{B}<1$. 

The physical significance of this non-trivial observation was
interpreted in the parton model \cite{Feynman:1969ej}, where the DIS 
process proceeds as shown in Fig.~\ref{Fig1}b, namely the electrons
scatter off nearly free electrically-charged pointlike particles 
called partons, with a cross-section that can be calculated in 
quantum electrodynamics (QED). The structure of the nucleon in 
DIS is described in terms of parton distribution functions (PDFs),
depicted by the green ellipse in Fig.~\ref{Fig1}b. 
(More precisely, PDFs are defined after squaring the amplitude in Fig.~\ref{Fig1} and summing over the complete set of states $X$.)
In modern
terminology, the PDFs in unpolarized DIS are denoted $f_1^a(x)$, 
with $a$ labelling the type of parton. More precisely,
$f_1^a(x)\,\ud x$ is the probability to find a parton of 
type $a$ in the initial state inside of a nucleon moving with 
nearly the speed of light 
(an appropriate picture in DIS where $x\approx x_B$) and carrying 
a fraction of the nucleon's momentum in the interval $[x,\;x+\ud x]$. 
It was soon realized that the electrically charged partons, 
identified with quarks and antiquarks, carry only half of the 
nucleon's momentum between them.

\subsection{Colored quarks and gluons, QCD, and confinement}
\label{sect-I.D}

The discovery of proton substructure and the development of the 
parton model were key to establishing quantum chromodynamics
(QCD) as the theory of the fundamental interaction between 
quarks carrying $N_c=3$ different color charges (and antiquarks
carrying the corresponding anticharges)
\cite{Gross:1973id,Politzer:1973fx,Fritzsch:1973pi}.
The color forces are mediated by the exchange of spin-1 gluons 
which also carry color charges (as opposed to electrically neutral 
photons which mediate interactions in QED). 
Evidence for the existence of gluons has been found in the study 
of $e^+e^-$ annihilation processes \cite{TASSO:1980lqw}. Being 
electrically neutral, the gluons are ``invisible'' in interactions 
with electrons, and account for the missing half of the 
proton momentum in DIS.

The QCD Lagrangian is given by
\be\label{Eq:Lagrangian}
      {\cal L} = \sum_q \overline{\psi}_q(i\slashed{D}+m_q)\psi_q
      - \tfrac{1}{4}\,F^2,
\ee 
where $\overline{\psi}_q$ and $\psi_q$ denote the quark and 
antiquark fields and $m_q$ denotes the current quark masses. 
The summation runs over the quark flavors 
$q\in\{u,\,d,\,s,\,c,\,b,\,t\}$. 
The covariant derivative is defined as 
$iD_\mu = i\partial_\mu+g A_\mu^cT^c$ and 
$F^2=F^c_{\mu\nu}F^{c\mu\nu}$ with 
$F^c_{\mu\nu}=\partial_\mu A^c_\nu-\partial_\nu A^c_\mu+g f^{cde}A^d_\mu A^e_\nu$.
Here $A_\mu^c$ are the gauge (gluon) fields and $T^c$ the generators in the fundamental representation 
of SU($N_c$) with $c\in\{1,\,\dots,\,N_c^2-1\}$ and $f^{cde}$ 
are the structure constants of the SU($N_c$) group. 
Non-abelian gauge theories like QCD are renormalizable
\cite{tHooft:1972tcz} with the coupling constant
$\alpha_s(\mu)=g(\mu)^2/(4\pi)$ depending on the 
renormalization scale $\mu$. 
When it comes to describing hadrons, the scale is $\mu \sim 1\,\rm{GeV}$ and $\alpha_s(\mu)$ is of order unity. The interaction is thus strong and the solution of (\ref{Eq:Lagrangian}) requires nonperturbative techniques.
However, in high-energy processes such as DIS, where the 
renormalization scale is identified with the hard scale of
the process, $\alpha_s(Q)$ decreases with increasing $Q$ reaching $\alpha_s(91\rm{~GeV})\approx 0.12$ at the scale of the $Z$-boson mass. 
 This property, known as asymptotic freedom, explains why 
quarks, antiquarks and gluons appear in such reactions as 
nearly free partons to a first approximation. The fact that 
free color charges are never observed in nature
gave rise to the confinement hypothesis, whose theoretical 
explanation is still an outstanding open question.

\subsection{Proton mass, spin and \boldmath $D$-term}

While the fundamental degrees of freedom and their interaction described in terms of the Lagrangian (\ref{Eq:Lagrangian}) are well-established, many questions remain open. For instance, 
the proton and neutron quantum numbers arise from combining 3 light quarks, $uud$ and $udd$, whose masses in the QCD Lagrangian (\ref{Eq:Lagrangian}) are explained by the Brout-Englert-Higgs mechanism \cite{Englert:2014zpa,Higgs:2014aqa}. The smallness of $m_u \sim 2\,{\rm MeV}/c^2$ and $m_d \sim 5\,{\rm MeV}/c^2$, however, gives rise to one of the central questions of QCD, namely how does the nucleon mass of 940 MeV$/c^2$ come about? 
(A wide-spread misconception is that $m_u+m_u+m_d \sim 9\,{\rm MeV}/c^2$ only explains about $1\,\%$ of the proton mass. This is incorrect, as in QCD the quark mass contribution is due to the operator $m_q\,\overline{\psi}_q\psi_q$ which 
includes virtual quark-antiquark pair contributions, leading to 
a much larger fraction (about 10-15\,\%) of the proton mass as will be discussed in Sec.~\ref{mass-spin}.)

Another central question concerns the proton spin. In a ``static'' quark model one would naively attribute the spin $\frac12$ of the nucleon to the spins of the quarks. In nature, due to the relatively light $u$- and $d$-quarks being confined within distances of ${\cal O}(1\,{\rm fm})$, Heisenberg's uncertainty principle implies an ultra-relativistic motion of the quarks. It must be expected that, e.g., the orbital motion of quarks has an important role in the spin budget of the nucleon. At the quantitative level, the nucleon spin decomposition is, however, still not known precisely \cite{Ji:2020ena}. 

The answers to these questions lie in the matrix elements of the energy-momentum tensor (EMT), 
an operator in quantum field theory of central importance 
that is associated with the invariance of the theory under 
spacetime translations. These matrix elements 
encode~key information including the mass 
and spin of a particle, the less well-known but equally 
fundamental $D$-term ($D$ stands for the German word {\it Druck} meaning pressure), as well as information about the 
distributions of energy, angular momentum, and various 
mechanical properties such as, e.g., internal forces inside the system. 
These properties are encoded in the gravitational form factors. In the standard model (plus gravity) the EMT couples to gravitons, so the direct way to measure its matrix elements would be graviton-proton scattering.  
Since the gravitational interaction between 
a proton and an electron is
(at currently achievable lab energies) $10^{-39}$
times weaker than their electromagnetic interaction, direct 
use of gravity to probe proton structure is impossible 
in electron-proton scattering, 
and in fact in any accelerator experiment in the foreseeable future.
However, we have learned how to apply indirect methods to acquire information about the EMT through studies of hard exclusive reactions.
The purpose of this Colloquium is to review the progress in theory, experiment, and interpretation of the EMT matrix elements. While the main focus is on the proton, also other hadrons will be discussed to provide a wider context and improve understanding.


\section{The energy-momentum tensor} \label{sec-II}

In this section, after reviewing the 
definition and properties of the EMT
in QCD, the gravitational form factors 
(GFFs) of the proton are
introduced. It is shown how GFFs can
be leveraged to elucidate the proton's 
mass and spin decompositions.

\subsection{Definition of the EMT operator}
\label{sec-II.A}
In QCD, the EMT 
$T^{\mu\nu}=\sum_qT^{\mu\nu}_q+T^{\mu\nu}_G$ can 
be decomposed into gauge-invariant quark and gluon
parts as 
\begin{equation}\label{EMTop}
    \begin{aligned} 
    T^{\mu\nu}_q
    &=\overline\psi_q\gamma^\mu\,iD^\nu\psi_q,\\
    T^{\mu\nu}_G
    &=-F^{c\mu\lambda}F^{c\nu}_{\phantom{c\nu}\lambda}
    +\tfrac{1}{4}\,g^{\mu\nu} F^2
    \end{aligned}  
\end{equation}
with $g_{\mu\nu}=\text{diag}(+1,-1,-1,-1)$ the 
Minkowski metric. In quantum field theory, the
expressions for the matrix elements of bare 
operators contain divergences and must be 
renormalized~\cite{tHooft:1972tcz}. Therefore,
each term in \eqref{EMTop} is understood as a 
renormalized operator defined at some 
renormalization scale $\mu$. 
The components of the EMT are interpreted in the 
same way as in the classical theory, namely $T^{00}$ 
is the energy density, $T^{0i}$ is the momentum density,
$T^{i0}$ is the energy flux, and $T^{ij}$ is the momentum
flux or stress tensor. 

Since the antisymmetric part $T^{[\mu\nu]}=\frac{1}{2}(T^{\mu\nu}-T^{\nu\mu})$ of \eqref{EMTop} can be written as a total divergence using the equations of motion, it does not contribute to the total four-momentum and angular momentum of the system. In the literature, one often
considers only the symmetric part $T^{\{\mu\nu\}}=\frac{1}{2}(T^{\mu\nu}+T^{\nu\mu})$, 
known as the Belinfante EMT~\cite{Belinfante:1962zz},
where the distinction between orbital angular momentum and spin is lost \cite{Leader:2013jra,Lorce:2017wkb}.

\subsection{Trace anomaly}
\label{Subsec:trace-anomaly}


The invariance of the classical Lagrangian of a theory under a
certain symmetry implies the existence of a conserved, so-called 
Noether, current \cite{Noether:1918zz}. For instance, 
the EMT is the Noether current associated with the 
invariance of a theory under space-time translations. 
If the classical symmetry is obeyed in quantum field theory
(as is the case for space-time translations) one obtains a
conservation law.

If a classical symmetry is spoiled by quantum effects, 
then one speaks of a ``quantum anomaly'' and there is no
associated conservation law. One important example is the 
trace anomaly 
(for another example see Sec.~\ref{Sec-4A-chiral-symmetry}): 
the QCD Lagrangian 
(\ref{Eq:Lagrangian}) is ``approximately'' invariant under 
scale transformations $x\mapsto x' = \lambda x$ with 
arbitrary $\lambda>0$. It is not an exact symmetry 
since the divergence of the corresponding Noether current 
does not vanish but is equal at the classical level to
$g_{\mu\nu}T^{\mu\nu}_\text{class}=\sum_qm_q\,\overline\psi_q\psi_q$.
In the light quark sector, due to the smallness of the 
up- and down-quark masses, one would nevertheless expect this 
to be a good approximate symmetry similarly to the isospin
symmetry encountered in Sec.~\ref{Sec-1A:magnetic-moment}.  
However, quantum corrections alter the trace of the EMT as
\cite{Collins:1976yq,Nielsen:1977sy}
\begin{equation}\label{EMTtrace}
    g_{\mu\nu}T^{\mu\nu}=\sum_q (1+\gamma_m)m_q\,\overline\psi_q\psi_q+\tfrac{\beta(g)}{2g}\,F^2,
\end{equation}
where $\gamma_m$ is the anomalous quark mass 
dimension and $\beta(g)=\partial g/\partial\ln\mu$ 
is the QCD beta function which describes how the coupling 
changes with the renormalization scale. As will be discussed 
later, the trace anomaly plays an important role for
the mass and mechanical properties of the proton. 
For more details, see \cite{Braun:2003rp} 
and \cite{Hatta:2018sqd,Tanaka:2018nae,Ahmed:2022adh}.

\subsection{Definition of the proton gravitational form factors}
\label{Subsec:def-GFFs}

The electromagnetic structure of the proton is encoded in the matrix elements of the electromagnetic current $\langle p',\vec s^{\,\prime}|J^{\mu}_{em}|p,\vec s\rangle$. Similarly, the matrix elements of the EMT operator $\la p^\prime,\vec s^{\,\prime}|T_a^{\mu\nu} |p,\vec s\rangle$ for quarks ($a=q$) and gluons ($a=G$) allow one to study 
the mass and spin decompositions, as well as the mechanical properties.

Thanks to Poincar\'e symmetry, these matrix elements can be 
written as~\cite{Kobzarev:1962wt,Pagels:1966zza,Ji:1996ek,Bakker:2004ib,Lorce:2022cle}
\begin{equation}\label{EMTparam}
\begin{aligned}
&   \la p^\prime,\vec s^{\,\prime}|  
    T_a^{\mu\nu} |p,\vec s\rangle
    =\overline u(p^\prime,\vec s^{\,\prime})\Bigg[A_a(t)\,\frac{P^\mu P^\nu}{M_N}\\
&   + D_a(t)\,
    \frac{\Delta^\mu\Delta^\nu-g^{\mu\nu}\Delta^2}{4M_N}+ \bar{C}_a(t)\,M_N\,g^{\mu\nu}\\
&   +
    J_a(t)\ \frac{P^{\{\mu}i\sigma^{\nu\}\lambda}
    \Delta_\lambda}{M_N}
    -S_a(t)\ \frac{P^{[\mu}i\sigma^{\nu]\lambda}
    \Delta_\lambda}{M_N}\Bigg]u(p,\vec s)
\end{aligned}    
\end{equation}
    with $P=(p'+p)/2$ and $\Delta=p'-p$ the symmetric kinematical variables, $u(p,\vec s)$ the usual free Dirac spinor, and $M_N$ the nucleon mass. The Lorentz-invariant functions $A_a(t)$, $D_a(t)$, $\bar C_a(t)$, $J_a(t)$ and $S_a(t)$ depend on the square of the four-momentum transfer $t=\Delta^2$. They are the EMT analogues of the more familiar electromagnetic FFs, and are accordingly called gravitational form factors (GFFs). In contrast to the electromagnetic FFs, these GFFs inherit also a renormalization
scale dependence from the associated operators, which 
is omitted in the notation for convenience. The \textit{total} 
GFFs $\sum_a A_a(t)$, $\sum_a D_a(t)$, $\sum_a\bar C_a(t)$ and $\sum_a J_a(t)$
are, however, renormalization scale independent~\cite{Nielsen:1977sy}.

On top of restricting the number of GFFs, Poincar\'e 
symmetry imposes additional constraints, namely
\ba
    A(0)&=&\sum_qA_q(0)+A_G(0)=1,  \label{Tconstraint}\\
    J(0)&=&\sum_qJ_q(0)+J_G(0)=\tfrac{1}{2}, \label{Lconstraint}\\
    \tfrac{1}{2}\Delta\Sigma &=& \sum_q S_q(0), \label{Lconstraint-2} \\
    \bar C(t)&=&\sum_q\bar C_q(t)+\bar C_G(t)=0, \label{cbar-constraint}
\ea
where (\ref{Tconstraint}) follows from translation
symmetry \cite{Ji:1997pf}, while~\eqref{Lconstraint} 
and~\eqref{Lconstraint-2} result from Lorentz symmetry 
\cite{Ji:1996ek,Bakker:2004ib}, with
$\tfrac{1}{2}\Delta\Sigma$ denoting the quark spin 
contribution to the nucleon spin. The constraint 
(\ref{cbar-constraint}), valid for any $t$,
follows from EMT conservation 
$\partial_\mu T^{\mu\nu}=0$. Interestingly, the
renormalization-scale invariant
quantity~\cite{Polyakov:1999gs} 
\begin{equation}\label{Eq:D-define}
    D\equiv D(0)=\sum_qD_q(0)+D_G(0),
\end{equation}
known as the $D$-term, is a global 
property of the proton (and, in fact, any hadron), 
whose value is not fixed by spacetime symmetries 
\cite{Polyakov:1999gs}. Its physical interpretation 
will be discussed in Sec.~\ref{interpretation}. 

Until recently, the only information about GFFs 
known from phenomenology was
$A_a(0)=\int_{-1}^1\ud x\, x\,f^a_1(x)$, corresponding 
to the fraction of proton momentum carried by the 
partons $a$ as inferred from DIS experiments, and
$S_q(0)=\frac{1}{2}\int_{-1}^1\ud x\, g^q_1(x)$, 
where $g^q_1(x)$ is the quark helicity 
distribution~\cite{Aidala:2012mv}.

\subsection{Decomposition of proton mass}
\label{mass-spin}

Just like the charge density is defined via a Fourier transform of the matrix elements of the electromagnetic current, the spatial distributions of energy and momentum read~\cite{Polyakov:2002yz,Polyakov:2018zvc,Lorce:2018egm}
\begin{equation}\label{Eq:static-EMT}
    \mathcal T^{\mu\nu}_a(\vec r)=\int\frac{\ud^3\Delta}{(2\pi)^32E}\,e^{-i\vec\Delta\cdot\vec r}\,\langle p'|T^{\mu\nu}_a|p\rangle
\end{equation}
in the so-called Breit frame defined by the conditions $\vec p^{\,\prime}=-\vec p=\vec\Delta/2$ and $p'^0=p^0=E=\sqrt{M_N^2+\vec\Delta^2/4}$. For ease of notation, the dependence on the nucleon polarization is omitted. Integrating over space, one obtains
\begin{equation}
    \int\ud^3r\,\mathcal T^{\mu\nu}_a(\vec r)=\frac{\langle p|T^{\mu\nu}_a|p\rangle}{2M_N}\bigg|_{\vec p=\vec 0}
\end{equation}
i.e.,~the matrix elements for the proton at rest. More explicitly, one finds 
\begin{equation}\label{EMTrest}
   \int\ud^3r\,\mathcal T^{\mu\nu}_a(\vec r)=\begin{pmatrix}U_a&0&0&0\\
    0&W_a&0&0\\ 0&0&W_a&0\\ 0&0&0&W_a\end{pmatrix}.
\end{equation}
The components $\mathcal T^{00}(\vec r)$ and $\frac{1}{3}\sum_i\mathcal T^{ii}(\vec r)$ represent the energy density and the isotropic pressure in the system, and so $U_a=\int\ud^3r\,\mathcal T^{00}_a(\vec r)=[A_a(0)+\bar C_a(0)]\,M_N$ and $W_a=\frac{1}{3}\sum_i\int\ud^3r\,\mathcal T^{ii}_a(\vec r)=-\bar C_a(0)M_N$ are respectively interpreted as the quark or gluon contributions to internal energy and pressure-volume work.

Since by definition $p^2=M_N^2$, the proton mass can be identified with the total energy in the rest frame
\begin{equation}\label{EnergySR}
    \sum_a U_a=M_N.
\end{equation}
Moreover, the proton being a bound state at mechanical equilibrium, the virial theorem says that the total pressure-volume work must vanish~\cite{Laue:1911lrk,Lorce:2017xzd,Lorce:2021xku}
\begin{equation}\label{sumrules}
    \sum_a W_a=0.
\end{equation}
These are two \textit{independent} sum rules underlying the various mass decompositions proposed in the literature, see~\cite{Lorce:2021xku} for a detailed review. To keep the following discussion as simple as possible, the standard $\overline{\text{MS}}$ scheme with the additional requirement that the trace anomaly arises purely from the gluonic sector is used in the following~\cite{Metz:2020vxd,Lorce:2021xku}.

Defining the quark mass contribution to the nucleon mass via
\begin{equation}
    M_m=\sum_q\sigma_q\equiv\frac{\langle p|\sum_q m_q\,\overline\psi_q\psi_q|p\rangle}{2M_N}\bigg|_{\vec p=\vec 0},
\end{equation}
one obtains a three-term mass decomposition directly from the energy sum rule~\eqref{EnergySR}
\begin{equation}\label{MassSR3term}
    M_N=\sum_qM_q+M_m+M_G,
\end{equation}
where $M_q=U_q-\sigma_q$ and $M_G=U_G$ can, respectively, be interpreted as the kinetic+potential energies of quarks and gluons~\cite{Rodini:2020pis,Metz:2020vxd}. Motivated by the fact that the traceless part of the gluon EMT can directly be accessed in high-energy experiments, a further of decomposition of the gluon energy
\begin{equation}\label{JiSR}
    M_G=\bar M_G+\tfrac{1}{4}M_A
\end{equation}
into the traceless part $\bar M_G=\frac{3}{4}(U_G+W_G)=\tfrac{3}{4}A_G(0)M_N$ and pure trace part $\frac{1}{4}M_A=\frac{1}{4}(U_G-3W_G)$ has been proposed in~\cite{Ji:1994av,Ji:1995sv,Ji:2021mtz}. Since at the classical level the gluon EMT is traceless, $\bar M_G$ was interpreted as the ``classical'' gluon energy and $\frac{1}{4}M_A$ with
\begin{equation}
  M_A=\frac{\langle p|\sum_q\gamma_mm_q\,\overline\psi_q\psi_q+\frac{\beta(g)}{2g}\,F^2|p\rangle}{2M_N}\bigg|_{\vec p=\vec 0}  
\end{equation}
as the ``quantum anomalous energy''. This interpretation is, however, not supported by a careful analysis in the $\overline{\text{MS}}$ scheme. Indeed, at the level of renormalized operators, it is the \textit{total} gluon energy density (and not its traceless part) that has the familiar form $T^{00}_G=\frac{1}{2}(\vec E^2+\vec B^2)$, ensuring that time translation symmetry remains \textit{exact} under renormalization~\cite{Nielsen:1977sy,Suzuki:2013gza,Tanaka:2018nae,Metz:2020vxd,Lorce:2021xku,Ahmed:2022adh,Tanaka:2022wzy}. A recent explicit one-loop calculation within the scalar diquark model~\cite{Amor-Quiroz:2023rke} confirms that, unlike the EMT trace, the total energy does not receive any intrinsic anomalous contribution.

Since mass is a Lorentz-invariant quantity, one sometimes prefers to start from the trace of the EMT
\begin{equation}
    \langle p|g_{\mu\nu}T^{\mu\nu}|p\rangle=2p^2=2M_N^2
\end{equation}
and then decompose it into quark and gluon contributions~\cite{Shifman:1978zn,Donoghue:1992dd,Hatta:2018sqd,Tanaka:2018nae}, leading to the sum rule
\begin{equation}\label{TraceSR}
    M_N=M_m+M_A.
\end{equation}
Current phenomenology~\cite{Hoferichter:2015hva} 
and Lattice QCD calculations~\cite{Alexandrou:2019brg} 
indicate that $M_m/M_N\approx 10\%$, suggesting that most of the proton mass comes 
from the trace anomaly (and hence from the gluons, 
since $\gamma_m$ is small). To clarify the actual meaning of this result, it has been noted in~\cite{Lorce:2017xzd} that the sum rule~\eqref{TraceSR} is equivalent to writing
\begin{equation}
    M_N=\sum_a\int\ud^3r\,g_{\mu\nu}\mathcal T^{\mu\nu}_a(\vec r)=\sum_a\left(U_a-3W_a\right).
\end{equation}
While the total pressure-volume work vanishes owing to the virial theorem~\eqref{sumrules}, it does nevertheless contribute to the \textit{separate} quark and gluon contributions to the EMT trace. Since $\sum_qU_q$ and $U_G$ turn out to be of the same 
order of magnitude, the smallness of $M_m$ 
relative to $M_A$ indicates in reality that $\sum_qW_q=-W_G>0$. In other words, the net quark force is repulsive and is exactly balanced by the net attractive gluon force.

Since the four-momentum (and hence the mass) of a system is defined via the $T^{0\mu}$ components of the EMT, it has been argued in~\cite{Lorce:2017xzd,Lorce:2021xku} that a genuine mass decomposition should in principle \textit{not} entail the components $T^{ii}$. In particular, the quantities $\bar M_g$ and $M_A$ involve the gluon pressure-volume work $W_g$, and hence do not have a clean interpretation as mass contributions. From this point of view, both~\eqref{JiSR} and~\eqref{TraceSR} should rather be regarded as mere sum rules mixing the genuine mass decomposition~\eqref{MassSR3term} with the virial theorem~\eqref{sumrules}.

\subsection{Decomposition of proton spin}

A similar discussion elucidates the proton spin 
decomposition. The total angular momentum (AM) operator 
is defined, in terms of the Belinfante (symmetric) EMT 
$T^{\mu\nu}_\text{Bel}=T^{\{\mu\nu\}}$, as 
\begin{equation}
\label{Eq:total-OAM-Bel}
    \mathcal J^i=\int\ud^3r\,\epsilon^{ijk}r^jT^{0k}_\text{Bel}.
\end{equation}
Because of the explicit factor of $r^j$, the expectation value of this operator in a momentum eigenstate turns out to be ill-defined. A proper treatment requires the use of wave 
packets and amounts to considering matrix elements 
with non-vanishing momentum 
transfer~\cite{Bakker:2004ib,Leader:2013jra}. 

For convenience, only the longitudinal AM 
(i.e.,~the component along the proton average momentum 
$\vec P=\frac{1}{2}(\vec p^{\,\prime}+\vec p)$ defining 
the $z$-direction) is considered here. The discussion 
about the transverse AM turns out to be much more complex 
because of its dependence on both $|\vec P|$ and the 
choice of origin, see e.g.~\cite{Lorce:2018zpf,Lorce:2021gxs} and references therein. From 
the splitting of the EMT in~\eqref{EMTop}, one finds that 
the quark and gluon contributions to the proton spin 
$\langle\mathcal J^z\rangle=\sum_qJ^z_q+J^z_G$ are 
given by~\cite{Ji:1996ek}
\begin{equation}
    J^z_a=J_a(0),
\end{equation}
for a proton polarized in the $z$-direction. 

Working instead with an asymmetric EMT, the quark AM 
operator can be further decomposed into orbital and 
intrinsic AM terms
\begin{equation}\label{Ja}
    \mathcal J^i_q=\int\ud^3r\,\epsilon^{ijk}r^jT^{0k}_q+\int\ud^3r\,\tfrac{1}{2}\overline\psi_q\gamma^i\gamma_5\psi_q.  
\end{equation}
Calculating the corresponding matrix elements, 
one then finds that $J^z_q=L^z_q+S^z_q$ with
\begin{equation}\label{quarkspinOAM}
\begin{aligned}
    L^z_q&=J_q(0)-S_q(0),\\
    \sum_qS^z_q&=\tfrac{1}{2}\Delta\Sigma.
\end{aligned}
\end{equation}
Combining the results~\eqref{Ja} and~\eqref{quarkspinOAM} 
with the fact that the proton is a spin-$\frac{1}{2}$ 
particle, one arrives at the constraints given 
in~\eqref{Lconstraint} and~\eqref{Lconstraint-2}. 

Since gluons are spin-$1$ particles, one may wonder 
whether the gluon AM could also be decomposed into 
orbital and intrinsic contributions. This can be done, 
but it requires non-local operators to preserve gauge 
invariance~\cite{Chen:2008ag,Hatta:2011ku,Lorce:2012rr,Lorce:2012ce,Leader:2013jra,Wakamatsu:2014zza}. 
One is then led to the canonical (or Jaffe-Manohar) 
spin decomposition~\cite{Jaffe:1989jz}, to be
distinguished from the one derived here from the 
local EMT~\eqref{EMTop} and known as the kinetic 
(or Ji) spin decomposition~\cite{Ji:1996ek}. Finally, it is possible to push this analysis further and study the spatial distribution of angular momentum~\cite{Lorce:2017wkb}.

\section{Measuring gravitational form factors} 
\label{sec:EMT-FF}

There is no {\sl direct} way to measure the proton GFFs, as it would require measurements of the graviton-proton  interaction~\cite{Kobzarev:1962wt,Pagels:1966zza}. More recent theoretical developments have shown, however, that the GFFs may be probed {\sl indirectly} in various exclusive processes. This is the subject of
this section.

\subsection{Deeply virtual Compton scattering (DVCS)}
\label{DVCS}

\begin{figure}[t]
\includegraphics[width=0.95\linewidth]{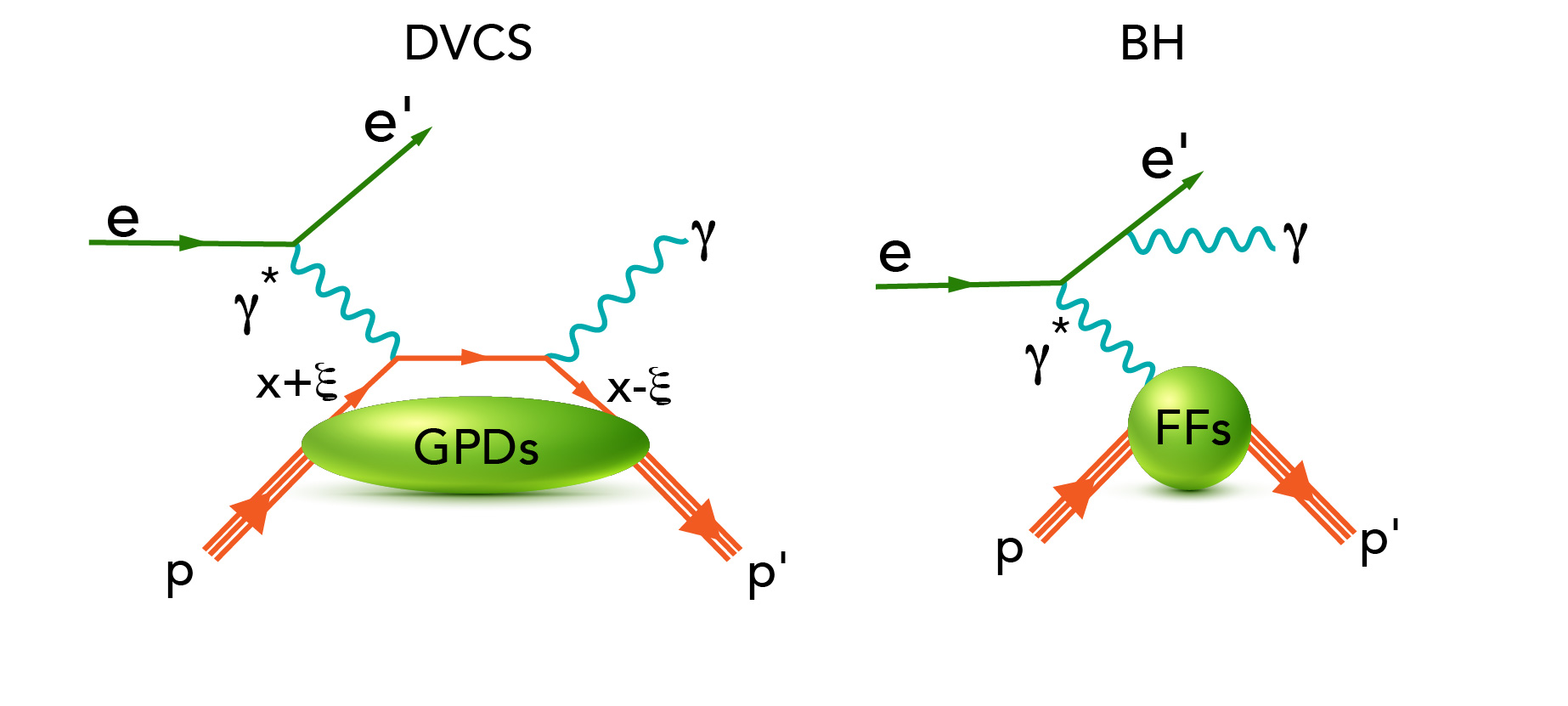}

\vspace{-10mm}\hspace{4mm}
{\footnotesize  (a) \hspace{30mm} (b)}
\vspace{-2mm}
\caption{\footnotesize\label{DVCS-BH} (a)  
QCD factorization of the DVCS amplitude.
The perturbatively calculable ``hard part'' 
is shown to lowest order in the strong coupling.
The nonperturbative ``soft part'' is described 
by the universal quark GPDs. 
(b) One of the QED diagrams for the 
amplitude of the Bethe-Heitler 
process, which has the same final state as DVCS 
and interferes with it. The  
Bethe-Heitler process is calculable requiring only the proton electromagnetic FFs as input.}
\end{figure}

In DVCS, the most explored process so far that accesses the GFFs,
high-energy charged leptons scatter off protons 
or nuclei by exchanging a deeply virtual photon,
producing a real photon in the final state
\cite{Muller:1994ses,Radyushkin:1996nd,Ji:1996nm}.
Similarly to DIS (see I.C), in the high-energy limit defined by $Q^2\to\infty$ and $P\cdot q\to\infty$ with $(-t)\ll Q^2$ and $P=(p'+p)/2$,
the process is described in QCD \cite{Collins:1998be} in terms of
the upper part of the handbag diagram shown in Fig.~\ref{DVCS-BH}a,
which can be calculated in perturbative QCD, and a lower part 
described in terms of generalized parton distributions 
(GPDs). GPDs are universal, i.e.,\ the same non-perturbative functions enter the description of different hard exclusive reactions. 

{GPDs are functions of $x$, $\xi$, and $t$. The new quantity $\xi\approx x_B/(2-x_B)$ in the high energy limit, called skewness, represents the longitudinal momentum transfer to the struck quark from the initial to final state (see Fig.~\ref{DVCS-BH}a). The variables $\xi$ and $t$ are observable in DVCS, while $x$ is not observable and enters the DVCS amplitude as an integration variable}.
GPDs encompass both PDFs and electromagnetic FFs 
discussed in Sec.~\ref{Sec-1:intro}.  
{For $p^\prime\to p$ implying $\xi\to 0$ and $t\to 0$, GPDs reduce to PDFs; integrating 
the GPDs over $x$ yields the electromagnetic FFs.}

GPDs parameterize the matrix elements of 
certain non-local operators which can be expanded in 
terms of a series of local operators with various $J^{PC}$ quantum numbers.
This includes operators with the quantum numbers of 
the graviton ($J=2$), and so part of the information 
about how the proton would interact with a graviton is
encoded within this tower. As the electromagnetic coupling to quarks is many orders of magnitude stronger than gravity, the DVCS process is an effective tool to probe the proton's gravitational properties.  Gluon GPDs are accessible in DVCS
only at higher orders in $\alpha_s$. 

The leading contribution to DVCS is described in terms of four GPDs. Two of them, namely $H_q(x,\xi,t)$ and $E_q(x,\xi,t)$, give access to the quark GFFs as follows
\begin{equation}\label{mellin-1}
\begin{aligned}
\int_{-1}^{1} \mathrm{d}x \, x H_q(x, \xi, t)  = A_q(t) + \xi^2 D_q(t),\\
\int_{-1}^{1} \mathrm{d}x \, x E_q(x, \xi, t) = B_q(t) -
\xi^2 D_q(t) ,
\end{aligned}
\end{equation}
where $B_q(t)=2J_q(t)-A_q(t)$
{is the quark contribution to the proton's anomalous gravitomagnetic moment. Analogous relations hold for gluons, and $B(0)=\sum_aB_a(0)$ vanishes due to Eqs.~\eqref{Tconstraint} and~\eqref{Lconstraint} \cite{Kobzarev:1962wt,Teryaev:1999su,Brodsky:2000ii,Lowdon:2017idv,Cotogno:2019xcl,Lorce:2019sbq}.}

The actual observables in DVCS are Compton form factors (CFFs)
{which are expressed by means of factorization formulae in terms of complex-valued convolution integrals given, at leading order $\alpha_s$, by}
\begin{multline} 
    \text{Re}{\mathcal H}(\xi,t) + i\, \text{Im}{\mathcal H}(\xi,t)  = \label{GPD-CFF} \\
    \sum_q e^2_q\int_{-1}^{1} dx \left[ \frac{1}{\xi-x-i\epsilon} -  \frac{1}{\xi+x-i\epsilon} \right] H_q(x,\xi,t),
\end{multline}
and similarly for the other GPDs. 
The CFFs 
are related to measurable quantities such as differential cross sections and beam and target polarization asymmetries.  

The DVCS cross section is typically very small.
Fortunately, DVCS interferes with the Bethe-Heitler 
process, see Fig.~\ref{DVCS-BH}b, which can be computed in QED given the proton's electromagnetic FFs, and has the 
same final state but with the final state photon emitted 
from the electron lines. 
The interference term projects out Im${\mathcal H(\xi, t)}$
when a spin-polarized
electron beam is employed, while Re${\mathcal H(\xi, t)}$ 
contributes dominantly to the unpolarized DVCS cross section,
and may be constrained through precise unpolarized cross 
section measurements.

The convolution integrals like (\ref{GPD-CFF}) cannot be inverted in a model-independent way
to yield GPDs \cite{Bertone:2021yyz}. 
However, with experimental information from other exclusive processes becoming available (to be discussed below), the GPDs may be further constrained. Presently, a model-independent extraction of the GPDs
and, via 
(\ref{mellin-1}), of the GFFs $A_q(t)$ and $J_q(t)$ 
is not possible. In the case of the GFF $D_q(t)$, however, the situation is more fortunate. In particular, the real and imaginary parts of ${\mathcal H(\xi,t)}$
are related by the fixed-$t$ dispersion relation 
\cite{Diehl:2007jb,Anikin:2007tx} 
\begin{multline}
 {\rm Re}{\mathcal H}(\xi,t) = \mathcal C_{\mathcal H}(t) \\+\frac{ 1}{\pi} \,{\text{P.V.}} \int_0^1 \text{d}\xi' \left[\frac{1}{\xi-\xi'}  -\frac{1}{\xi+\xi'}\right] {\rm Im}{\mathcal H}(\xi',t), \label{DR}
\end{multline}
where $\text{P.V.}$ denotes the Cauchy's principal value of the integral. This expression contains a real subtraction term $\mathcal C_{\mathcal H}(t)$ given by 
\begin{equation}
    \mathcal C_{\mathcal H}(t)=2\sum_qe^2_q\int_{-1}^1\ud z\,\frac{D^q_\text{term}(z,t)}{1-z},
\end{equation}
where $D^q_\text{term}(z,t)$, originally introduced in
\cite{Polyakov:1999gs}
and further elucidated in
\cite{Teryaev:2001qm},
has the expansion \cite{Goeke:2001tz}
\begin{equation}\label{Gegenbauer}
    D^q_\text{term}(z,t)=(1-z^2)\sum_{\text{odd}\,n}d^q_n(t)\,C^{3/2}_n(z)
\end{equation}
with $C^\alpha_n(z)$ the Gegenbauer polynomials
which diagon\-alize the leading-order evolution equations (the renormalization scale dependence is not indicated throughout this work).
In the limit of renormalization scale $\mu\to\infty$,
all $d^q_n(t)$ go to zero except $d^q_1(t)$, which is 
related to the GFF $D_q(t)$ as follows
\begin{equation}\label{Dtermint}
    D_q(t) = \frac{4}{5}\, d^q_1(t) =
    \int_{-1}^1\ud z\,z\,D^q_\text{term}(z,t) \,.
\end{equation}
Thus, extracting information on 
Im${\mathcal H(\xi,t)}$ and Re${\mathcal H(\xi,t)}$ and their scale dependence from experimental data provides access to the GFF $D_q(t)$.

\subsection{DVCS with positron and electron beams}

When data with both positron and electron beams are available, 
it is possible to measure the beam charge asymmetry $A_C$ defined as the difference in the $ep\to ep\gamma$ cross section when measured with an electron beam and measured with a positron beam, divided by their sum
\begin{eqnarray}
A_C = \frac{\sigma_{}^{e^-}-\sigma_{}^{e^+}}{\sigma_{}^{e^-}+\sigma_{}^{e^+}}. 
\end{eqnarray} 
The numerator of $A_C$ is given
by the real part of the DVCS and Bethe-Heitler interference term providing the cleanest access to Re$\mathcal{H}$ \cite{Kivel:2000fg,Belitsky:2001ns}. 
In contrast to this, in DVCS measured with electrons (or positrons)
alone, additional theoretical assumptions in the CFF extraction procedure 
are unavoidable~\cite{CLAS:2021gwi}. 

\begin{figure}[t]
\includegraphics[width=1.0\columnwidth]{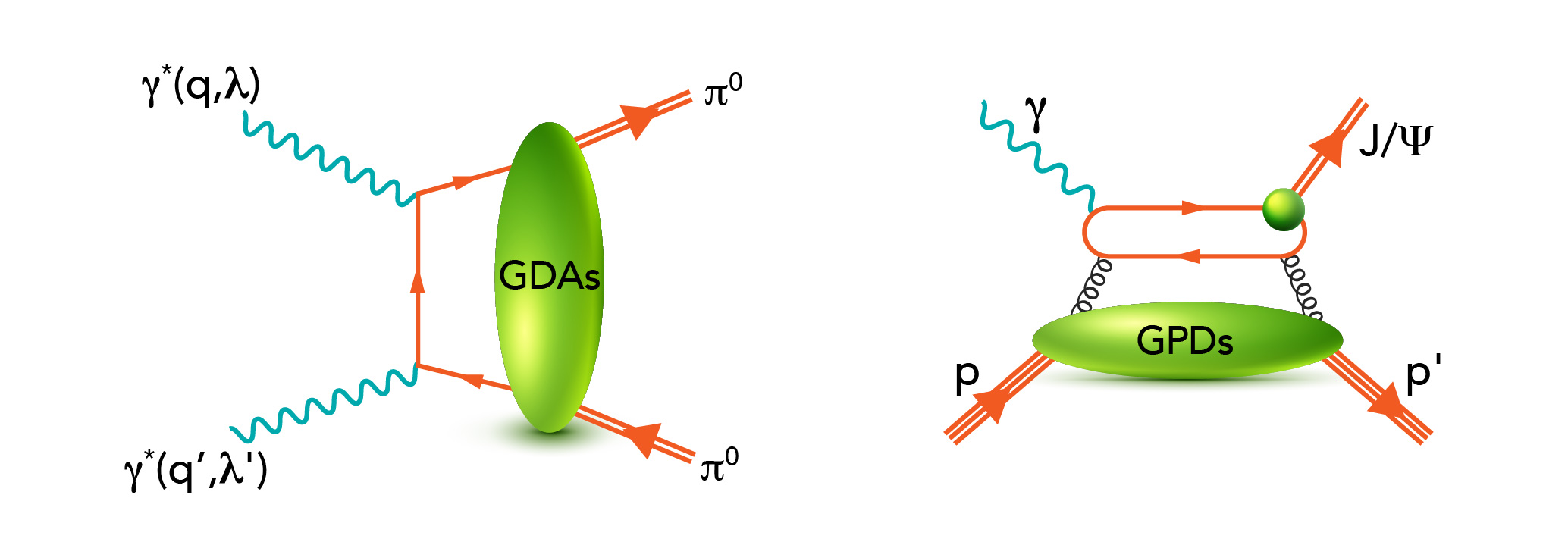}

\vspace{-3mm}
{\footnotesize  (a) \hspace{35mm} (b)}

\vspace{-2mm}
\caption{\footnotesize  
(a) The process $\gamma\gamma^\ast\to\pi^0\pi^0$ 
is described in terms of generalized distribution 
amplitudes, 
which provide access to GFFs in 
the time-like region $t>0$.
(b) Threshold $J/\Psi$ photo-production on the proton. 
This process is sensitive to the gluon GPDs.  
}
\label{Jpsi-gamma-gamma-pi-pi}
\end{figure}

\vspace{-4mm}

\subsection{\boldmath $\gamma\gamma^\ast\to\pi^0\pi^0$}
\label{Subsec:Jpsi-pi0-GDA}

\vspace{-2mm}

The process $\gamma\gamma^\ast\to\pi^0\pi^0$ shown in 
Fig.~\ref{Jpsi-gamma-gamma-pi-pi}a
can be studied, e.g., at electron-positron
colliders, and is described in terms of generalized 
distribution amplitudes which correspond to GPDs 
continued analytically from the $t$- to the $s$-channel~\cite{Muller:1994ses,Diehl:1998dk}.
In this way, one can access information
on GFFs in the time-like region where $t>0$~\cite{Kumano:2017lhr,Lorce:2022tiq}.
This process provides a unique opportunity to study 
the structure of unstable hadrons like pions that are not available
as targets.

\subsection{Time-like Compton scattering and double DVCS} 

Several other processes provide complementary 
information about the nucleon GFFs. One of them
is time-like Compton scattering (TCS), $\gamma p \to p'\gamma^\ast$,
where the final state virtual photon produces 
an $e^+e^-$ pair \cite{Berger:2001xd,Pire:2011st,CLAS:2021lky}.
In TCS, Im$\mathcal{H}$ can be accessed through the polarized beam 
spin asymmetry and Re$\mathcal{H}$ through a forward-backward
asymmetry of the final-state $e^+e^-$ pair
in its centre-of-mass frame.

\begin{figure}[ht]
\vspace{-2.5mm}
    \centering    \includegraphics[width=1.0\columnwidth]{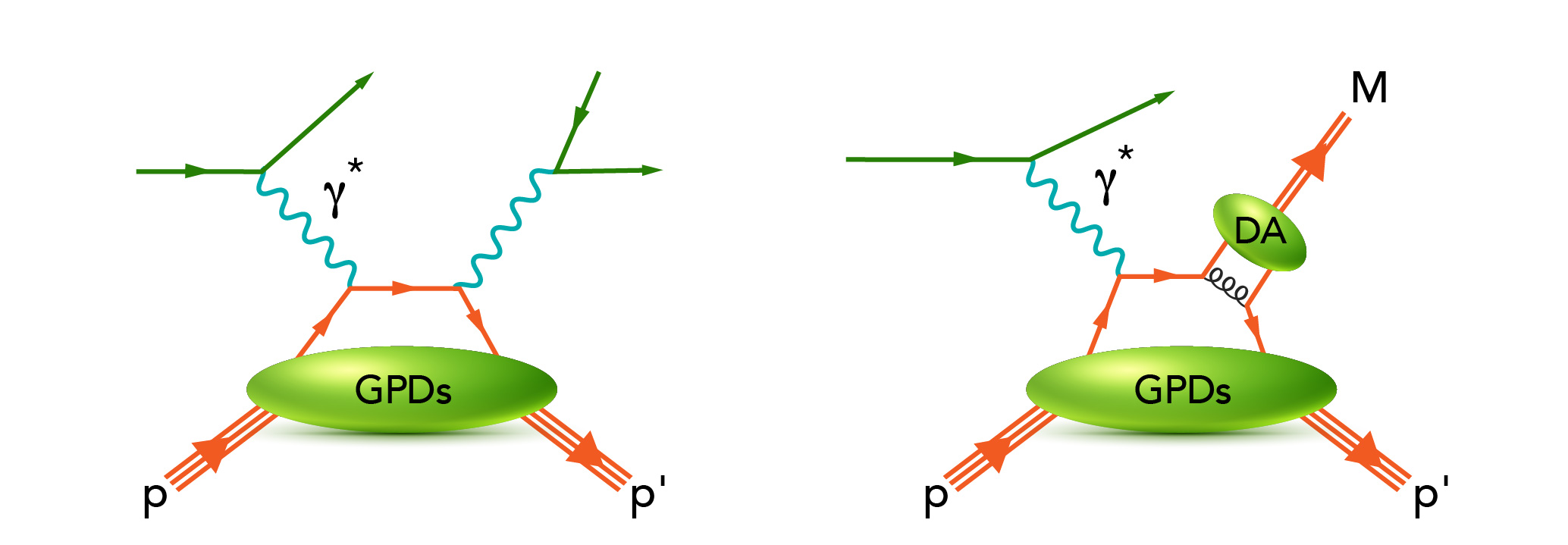}
{\footnotesize  (a) \hspace{37mm} (b)}
\vspace{-2mm}
    \caption{\footnotesize The leading double DVCS diagram 
    (a) and one of the leading diagrams for deeply
    virtual meson production (b).
    The ellipse where the meson M is produced
    is the nonperturbative distribution amplitude DA. 
    \label{fig:ddvcs-dvmp}}
\end{figure}

The double DVCS process~\cite{Belitsky:2002tf,Guidal:2002kt} displayed in 
Fig.~\ref{fig:ddvcs-dvmp}a may also play an important role at future facilities. 
It is a variant of DVCS 
with the final-state time-like photon converting into a $e^+e^-$ or $\mu^+\mu^-$  pair.
While in DVCS the GPDs are sampled along the lines
$x=\pm\xi$ in the convolution integrals (\ref{GPD-CFF}), 
this constraint is relaxed in double DVCS due 
to the variable invariant mass of the  
lepton pair. This 
is an advantage of this process, and 
will be of importance for less model-dependent global extractions of GPDs. 

\subsection{Meson production}
\label{meson production}

Deeply virtual meson production 
\cite{Collins:1996fb}
is another process sensitive to 
GPDs, see Fig.~\ref{fig:ddvcs-dvmp}b.
Production of different vector mesons provides 
sensitivity to GPDs of different quark flavors
which is an advantage over DVCS. 
However, this process is more difficult to analyze
than DVCS since gluons contribute on the same
footing as quarks (Fig.~\ref{fig:ddvcs-dvmp}b
only shows a quark diagram) and one 
in general expects larger power corrections.
{Also the process of heavy vector quarkonium photoproduction was shown to factorize in the heavy quark limit at one-loop order in perturbative QCD \cite{Ivanov:2004vd}.}

Exclusive $J/\Psi$ photo-production at threshold,
is expected to be sensitive to gluon GFFs \cite{Kharzeev:1995ij,Kharzeev:2021qkd} and more generally, as depicted in Fig.~\ref{Jpsi-gamma-gamma-pi-pi}b, to gluon GPDs \cite{Hatta:2018ina,Guo:2021ibg}, 
which in DVCS are accessible only at higher orders in $\alpha_s$.

Gluon GFFs have recently been extracted from this process by \cite{Duran:2022xag}, but the link with the physical observables is not direct and requires approximations \cite{Sun:2021gmi,Sun:2021pyw}, similarly to DVCS. $J/\Psi$ photoproduction can also be studied with quasi-real photons of virtualities as low as $Q^2 \lesssim 0.1\,{\rm GeV}^2$ emitted by electrons, 
together with electroproduction and DVCS.

Finally, a new class of hard scattering processes with multi-particle final states
has recently emerged~\cite{Qiu:2022bpq,Pedrak:2020mfm,Grocholski:2022rqj,
Ivanov:2002jj,Boussarie:2016qop,Duplancic:2018bum}. 
Those reactions are theoretically appealing, but measuring them is challenging.

{The relatively recent progress reviewed here paved the way to the exciting and even more recent experimental developments, which will be reviewed in Sec.~\ref{experiments} with a focus on DVCS and TCS.
Before continuing, the next section is devoted to the theory of GFFs, whose history is equally interesting and began much earlier.}

\section{Theoretical Results}
\label{Sec-5:theory}

GFFs were introduced by \cite{Kobzarev:1962wt}
who considered spin-0 and spin-$\frac12$
particles and parity-violating weak effects 
(not discussed here), proved the
vanishing of proton's anomalous gravitomagnetic moment $B(0)=0$,
and showed that one would need energies around the 
Planck scale to measure GFFs in gravitational
interactions. This section presents an overview of GFFs from the theory perspective with 
particular focus on $D(t)$, the least known of the total GFFs.
Despite the focus on the proton, it will be insightful to mention other hadrons for comparison when appropriate.

\subsection{\boldmath Chiral symmetry and the $D$-term of the pion}
\label{Sec-4A-chiral-symmetry}

GFFs received little attention from the community
until it was realized that matrix elements such as 
$\langle\pi,\pi|T^{\mu\nu}|0\rangle$
enter the QCD description of hadronic decays of
charmonia \cite{Novikov:1980fa,Voloshin:1980zf} 
or the decay of a hypothetical light Higgs
boson, an idea entertained in the early 1990s
when the possibility of a light Higgs was not yet
experimentally excluded \cite{Donoghue:1990xh}.
These matrix elements are related to pion GFFs
in the timelike region $t>0$. 

In general, hadronic EMT matrix elements cannot 
be computed analytically in QCD, but the pion is a notable 
exception. The QCD Lagrangian~\eqref{Eq:Lagrangian} 
exhibits a classical symmetry under global left- 
and right-handed rotations in the flavor space of
up, down and strange quarks. This symmetry is approximate
due to the small but non-zero quark masses $m_q$. 
If this symmetry were realized in nature, then 
{for example the nucleon state N(940) 
(here N stands for a state with nucleon isopin quantum number
and the number in the brackets is the rounded mass of
the state in ${\rm GeV}/c^2$) 
with the spin-parity quantum numbers $J^P={\frac12}^+$
should have the same mass as its negative-parity partner
$N(1535)$ with $J^P={\frac12}^-$ modulo small
corrections due to the small $m_q$.}
However, the latter is 
almost 600 MeV$/c^2$ heavier than the nucleon, an 
effect that cannot be attributed to current 
quark mass effects. The phenomenon that a 
symmetry of the Lagrangian is not realized in
the particle spectrum is known as spontaneous 
symmetry breaking \cite{Nambu:1961tp,Nambu:1961fr}. 
It is accompanied by the  emergence of massless Goldstone
bosons, corresponding in QCD to pions, kaons, and $\eta$-mesons,
which are not massless but are very light 
compared to other hadrons. 

In theoretical calculations, chiral symmetry is a powerful 
tool allowing one to evaluate the matrix elements of 
Goldstone bosons in the chiral limit (and for $t\to0$).
In this way, one obtains for the pion (and kaon and $\eta$) $D$-term \cite{Novikov:1980fa}
\be
       \lim\limits_{m_\pi\to 0} D_\pi = -1.
\ee
Deviations from the chiral limit 
are systematically calculable in chiral
perturbation theory \cite{Donoghue:1991qv}
and are expected to be small for pions and 
more sizable for kaons and the
$\eta$-meson \cite{Hudson:2017xug}.
The relation between the stability of the pion and spontaneous chiral symmetry breaking was discussed by~\cite{Son:2014sna}, and the
gravitational interactions of 
Goldstone bosons were studied by 
\cite{Voloshin:1982eb,Leutwyler:1989tn}.
{
For hadrons other than pions, the techniques based on the chiral limit of QCD cannot predict the $D$-term, but they can still be explored to provide insights on some properties of $D(t)$, as will be discussed in Sec.~\ref{Sec-4C-limits-in-QCD}. }

\subsection{GFFs in model studies}
\label{Sec-4B-in-models}

Interest in GFFs was once again renewed 
after it was shown that they can be
inferred from hard-exclusive reactions 
via GPDs and play a key role for the
understanding of the mass and spin 
structure of the proton, see 
Sec.~\ref{sec-II}, and further stimulated
by their interpretation in terms of forces
inside hadrons \cite{Polyakov:2002yz}.
The first model study of proton GFFs was
presented by \cite{Ji:1997gm} in the bag model, followed by works in
the chiral quark-soliton model  \cite{Petrov:1998kf,Schweitzer:2002nm,Ossmann:2004bp,Goeke:2007fp,Goeke:2007fq,Wakamatsu:2007uc,Kim:2021jjf}
and Skyrme models \cite{Cebulla:2007ei,Jung:2013bya,Perevalova:2016dln}. 

Extensive GFF model studies for the nucleon 
and other hadrons were presented in 
light-front constituent quark models
    \cite{Pasquini:2007xz,Sun:2020wfo},
diquark approaches \cite{Hwang:2007tb,Kumar:2017dbf,Chakrabarti:2020kdc,Choudhary:2022den,Fu:2022rkn,Amor-Quiroz:2023rke}, 
holographic AdS/QCD models 
\cite{Abidin:2008hn,Abidin:2009hr,Brodsky:2008pf,Chakrabarti:2015lba,Mondal:2015fok,Mondal:2016xsm,Mamo:2019mka,Mamo:2021krl,Mamo:2022eui,Fujita:2022jus},
a large-$N_c$ bag model 
    \cite{Neubelt:2019sou,Lorce:2022cle},
a cloudy bag model 
    \cite{Owa:2021hnj},
light-cone QCD sum rules
\cite{Anikin:2019kwi,Azizi:2019ytx,Aliev:2020aih,Azizi:2020jog,Ozdem:2020ieh},
the Nambu--Jona-Lasinio model
    \cite{Freese:2019bhb},
chiral quark-soliton model with strange
and heavier quarks
    \cite{Kim:2020nug,Won:2022cyy,Ghim:2022zob}, 
a dual model with complex Regge trajectories
    \cite{Fiore:2021wuj}
and in an instant-form relativistic 
impulse approximation approach
    \cite{Krutov:2020ewr,Krutov:2022zgg}.
Algebraic GPD Ans\"atze were used to shed
light on pion and kaon GFFs 
    \cite{Raya:2021zrz}
and toy models 
    \cite{Kim:2022wkc}
as well as light-cone convolution models 
    \cite{Freese:2022yur}
were used to study the deuteron GFFs.

The $D$-terms of nuclei were studied in
the liquid-drop model 
    \cite{Polyakov:2002yz},
revealing that for nuclei $D(0)\propto A^{7/3}$ 
grows strongly with mass number $A$. Studies 
in the Walecka model 
    \cite{Guzey:2005ba}
support this prediction which can 
be tested in DVCS experiments with
nuclear targets. Different results were
obtained in a non-relativistic nuclear 
spectral function approach 
    \cite{Liuti:2005qj}. 
Nuclear GFFs were also investigated
in Skyrme model frameworks
\cite{Kim:2012ts,Jung:2014jja,Kim:2022syw,GarciaMartin-Caro:2023klo}.

The GFFs for a constituent quark were studied
in a light-front Hamiltonian approach
    \cite{More:2021stk,More:2023pcy} 
which, after rescaling and regularization 
of infrared divergences, reproduces QED 
results for an electron
    \cite{Metz:2021lqv,Freese:2022jlu}. 
GFFs of the photon in QED were studied in
\cite{Friot:2006mm,Gabdrakhmanov:2012aa,Polyakov:2019lbq,Freese:2022ibw}.
An insightful model for composite 
particles is the $Q$-ball system where stable, metastable, unstable states 
were investigated, showing that, among all studied particle
properties, $D(0)$ is most sensitive to 
details of the dynamics
    \cite{Mai:2012yc,Mai:2012cx,Cantara:2015sna}.
Remarkably, the same conclusions were obtained in 
the bag model where, e.g., for the $N^{\rm th}$
highly excited nucleon state the mass increases as $M\propto N^3$ 
whereas $D(0)\propto N^8$ grows much more
strongly with $N$ \cite{Neubelt:2019sou}. 

\subsection{\boldmath Limits in QCD and 
dispersion relations}
\label{Sec-4C-limits-in-QCD}

Model-independent results for GFFs can be obtained 
in certain limiting situations in QCD, e.g., when the 
number of colors $N_c\to\infty$ or when $|t|$
becomes very small or very large, and through the use
of dispersion relation methods. These methods
are complementary to the nonperturbative lattice QCD methods which are reviewed in the next
section.

In the large-$N_c$ limit of QCD, baryons
are described as solitons of mesonic fields
\cite{Witten:1979kh}. Large-$N_c$ QCD has not been
solved (in 3+1 dimensions) and the soliton field
is not known (though it can be modelled).
Nontrivial results can, however, be derived 
based on the known symmetries of the
large-$N_c$ soliton field which are generally
well-supported in nature \cite{Dashen:1993jt}
despite $N_c=3$.
The relations of the GFFs of the nucleon and
$\Delta$ were studied in the large-$N_c$ limit 
of QCD in \cite{Panteleeva:2020ejw}.
The GFFs of the $\Delta$ are difficult to
measure, but such relations can be tested, e.g., in 
soliton models like the chiral quark-soliton model
or Skyrme model (mentioned in the previous subsection)
or in lattice QCD, discussed in the next section.

At small $|t|$, one can use chiral perturbation theory,
where one writes down an effective Lagrangian in terms of hadronic degrees of freedom with the most general interactions allowed by chiral symmetry, and free parameters which can be inferred from comparison of observable quantities with experiment.
A pioneering study to lowest order in chiral perturbation
theory was presented in \cite{Belitsky:2002jp} and 
studies at next-to-leading order \cite{Diehl:2006ya}
have been completed in
\cite{Alharazin:2020yjv}. In this way, one can obtain valuable model-independent information on the
$t$-dependence of GFFs for small $t$. For instance, for the nucleon 
the slope of $D(t)$ at $t=0$ diverges in the chiral
limit as
\be
      \frac{\ud}{\ud t}\,D(t)\bigg|_{t=0} = -
      \frac{g_A^2 M_N}{40\pi f_\pi^2 m_\pi} + \dots,
\ee
where $g_A=1.26$ is the isovector axial constant, 
$f_\pi = 93\,\rm MeV$ is the pion decay constant,
$m_\pi$ is the pion mass, and the dots indicate
(finite) higher-order chiral corrections. Such
results are reproduced in chiral soliton models
\cite{Goeke:2007fp,Cebulla:2007ei}.
The value of the $D$-term cannot be determined 
exactly in chiral perturbation theory for hadrons
other than Goldstone bosons. It is, however, 
possible to derive an upper bound, e.g., 
for the nucleon $D/M_N \le -(1.1\pm 0.1)~\rm GeV^{-1}$
in the chiral limit \cite{Gegelia:2021wnj}.
The GFFs of the $\rho$-meson
\cite{Epelbaum:2021ahi} and $\Delta$-resonance
\cite{Alharazin:2022wjj} have also been studied in chiral
perturbation theory.

Model-independent results for GFFs can also
be derived for asymptotically large momentum
transfers using power counting and perturbative
QCD methods
\cite{Tanaka:2018wea,Tong:2021ctu,Tong:2022zax}.
For instance, the proton GFFs $A_a(t)$ 
for quarks and gluons behave like $1/t^2$
at large $(-t)\to\infty$.
Since QCD factorization of hard exclusive processes
requires $(-t)\ll Q^2$ and $Q^2$ is in practice often
not large in current experimental settings, such results provide important theoretical guidelines 
to extrapolate to larger $|t|$. However,
based on experience
with analogous perturbative QCD predictions for 
the electromagnetic pion form factor, see 
e.g.\ \cite{Horn:2016rip} for a review, it is 
difficult to anticipate how large the momentum
transfer $t$ must be for a form factor to reach 
the asymptotic regime.

A theoretical study of the quark contribution to the nucleon GFF $D_q(t)$ in the range $0 < (-t) < 1\,{\rm GeV}^2$ was presented in \cite{Pasquini:2014vua} based on dispersion theory methods which rely on general principles like relativity, causality and unitarity. This approach does not require modelling other than making use of available information on pion-nucleon partial-wave helicity amplitudes and relying on mild assumptions like the saturation of the $t$-channel unitarity relation in terms of the two-pion intermediate states or input pion PDF parametrizations. 

\subsection{Lattice QCD}

Complementing the insights gained from models of proton and nuclear structure, numerical lattice QCD calculations give direct and controllable QCD predictions for matrix elements of the EMT operator. 
In particular, lattice QCD is the only known systematically improvable approach to computing observables in QCD in the low-energy (non-perturbative) regime. The approach proceeds via a discretisation of the QCD Lagrangian (\ref{Eq:Lagrangian}) onto a Euclidean space-time lattice, with a finite lattice spacing which is not physical but acts as a method of regularisation of the theory. Calculations then proceed via Monte-Carlo integration of the high-dimensional discretised path-integral; continuum QCD results are recovered in the limit of vanishing lattice discretisation scale, infinite lattice volume, and precise matching of the bare quark masses to reproduce simple physical observables. By this approach, matrix elements of local operators, such as the separated quark and gluon components of the EMT in proton or nuclear states, may be computed directly.

\begin{figure}[ht!]
\includegraphics[width=1.0\columnwidth]{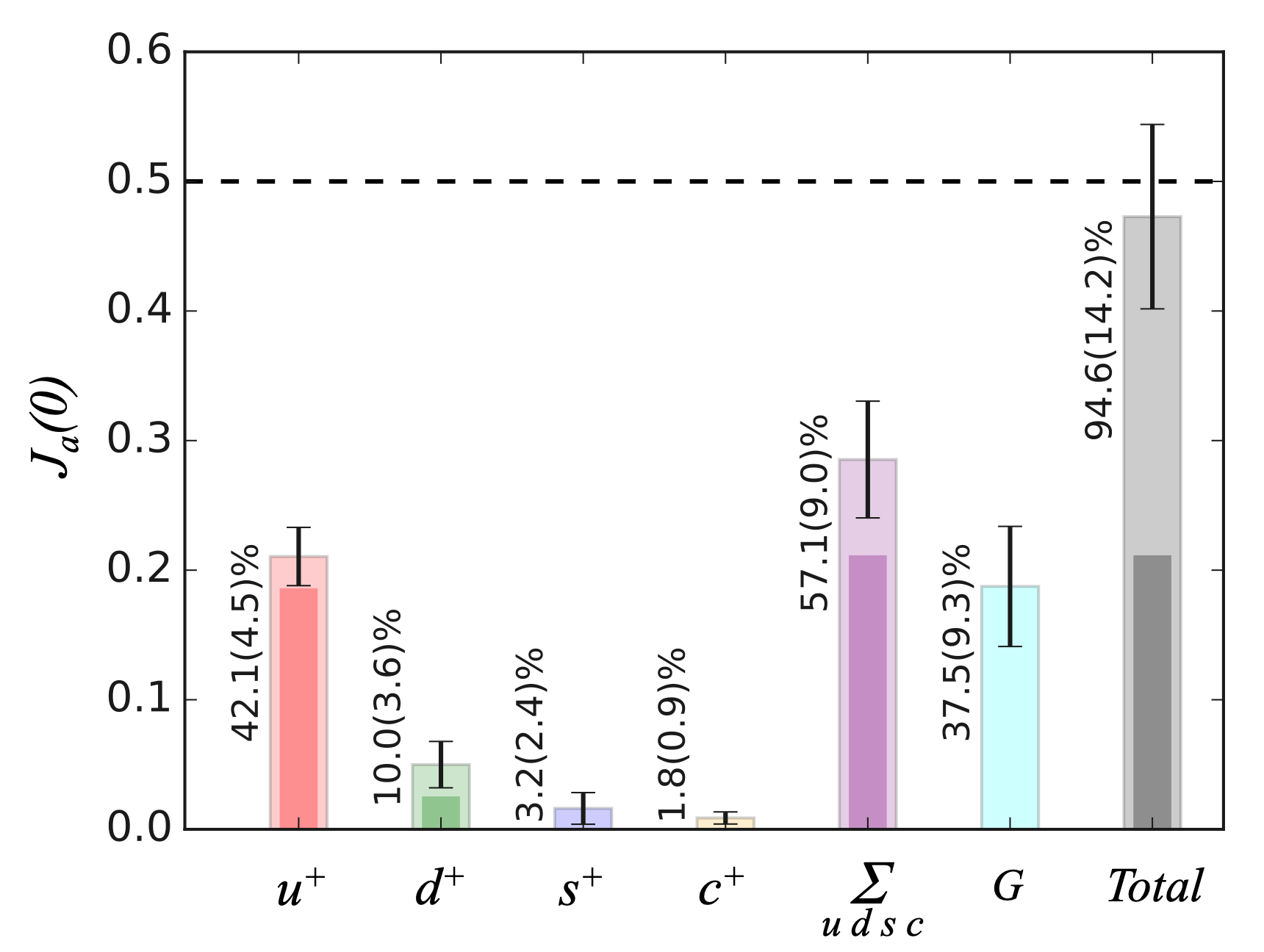}
\caption{\footnotesize Proton spin decomposition computed in lattice QCD in ~\cite{Alexandrou:2020sml}, given in the $\overline{\text{MS}}$ scheme at 2~GeV. Each component includes the contribution of both the quarks and antiquarks ($q^+=q+\overline{q}$); outer/light (inner/dark) shaded bars denote the total (purely connected) contributions.}
\label{fig:spindecomp}
\end{figure}

In the current era of precision lattice QCD calculations of proton structure, particular efforts have been made to determine the complete decomposition of the proton's spin and momentum into individual quark and gluon contributions with high precision and systematic control. For example, recent lattice QCD studies have isolated all angular momentum components in the kinetic (or Ji) decomposition~\cite{Alexandrou:2020sml,Wang:2021vqy}, with $\approx 10\%$ uncertainty in the total quark and gluon contributions; the results from one collaboration are shown in Fig.~\ref{fig:spindecomp}. This example illustrates the complementarity between theory and experiment in this area; flavour separation in lattice QCD calculations is in principle more straightforward, although some contributions, such as those from gluons or arising from ``disconnected'' contributions, e.g. strange and charm quarks in the proton, are difficult to compute because of signal-to-noise challenges. Computing the gluon spin and orbital angular momentum in the Jaffe-Manohar decomposition introduces additional challenges to the lattice QCD approach, but first results have been achieved based on constructions using both local and non-local operators~\cite{Yang:2016plb,Engelhardt:2020qtg}. 

\begin{figure}[hbpt]

\includegraphics[width=1.03\columnwidth]{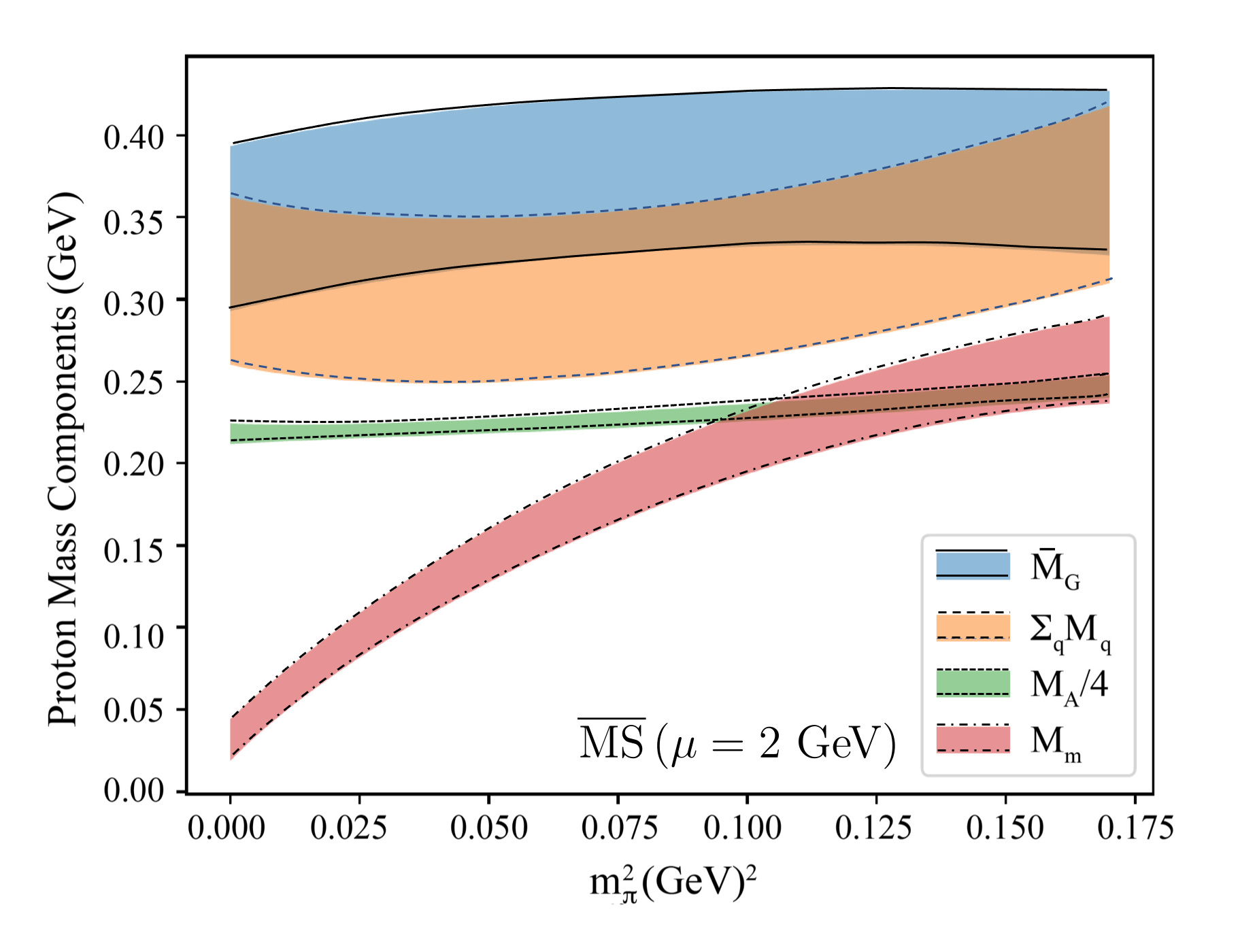}

\caption{\footnotesize Ji's mass decomposition (i.e.~combination of~\eqref{MassSR3term} and~\eqref{JiSR}) for a proton computed in lattice QCD in~\cite{Yang:2018nqn} at a scale $\mu=2$ GeV, as a function of the pion mass.}
\label{fig:massdecomp}

\vspace{5mm}

\includegraphics[width=1.01\columnwidth]{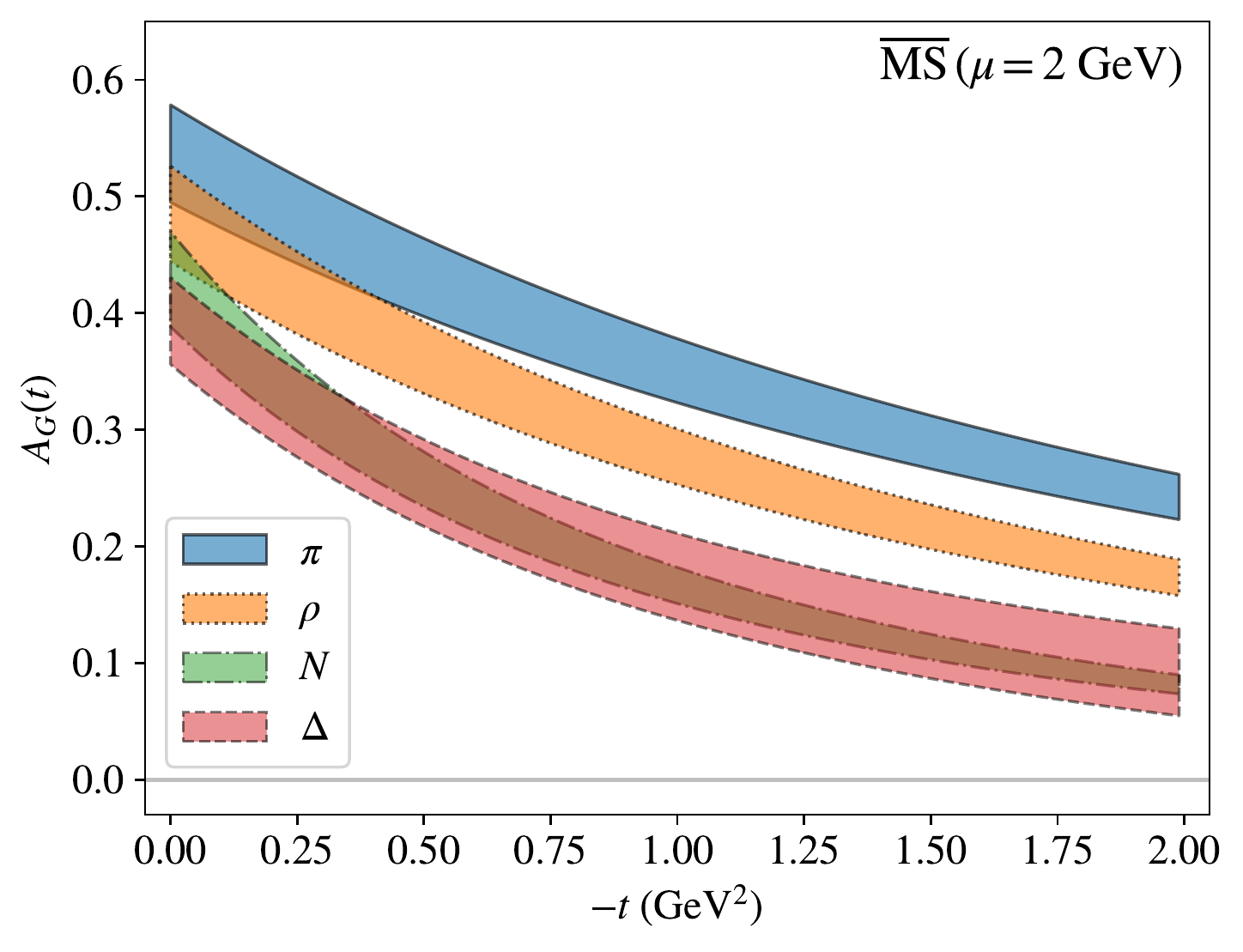}
\caption{\footnotesize $A_G(t)$ GFF for various hadrons from \cite{Pefkou:2021fni}, with quark masses corresponding to a larger-than-physical value of the pion mass of 450~MeV. }
\label{fig:AgGFFLattice}
\end{figure}

In the same vein, precise decompositions of the quark and gluon contributions to the proton's momentum, which are related to the mass decomposition, have been achieved with complete systematic control in the same computational frameworks that yielded the spin decomposition~\cite{Alexandrou:2020sml,Wang:2021vqy}. Contributions from the trace anomaly to the proton's mass decomposition are more difficult to compute directly with systematic control, but have been constrained using the trace sum rule~\eqref{TraceSR}; Fig.~\ref{fig:massdecomp} shows the first insight from lattice QCD into the pion mass (or quark mass) dependence of the proton's mass decomposition~\cite{Yang:2018nqn}. It is particularly notable that while the quark scalar condensate contribution varies rapidly with quark mass, the other contributions, including that of the trace anomaly, remain approximately constant. 

While local matrix elements in nuclear states can in principle be computed in lattice QCD in the same way as in the proton state, such calculations face significant practical and computational challenges, in particular compounding factorial and exponential growth in computational cost with the atomic number of the nuclear state. To date, a single first-principles calculation of 
isovector quark momentum fraction $A_{u-d}(0)$ in $^3\text{He}$~\cite{Detmold:2020snb} has been achieved; despite significant systematic uncertainties, including the result into global fits of experimental lepton-nucleus scattering data yields improved constraints on the nuclear parton distributions. Over the coming decade, it can be anticipated that the control and precision achieved in first-principles calculations of simple aspects of the gravitational structure of the proton will be extended to nuclear states.

\begin{figure}[tp!]
\includegraphics[width=1.0\columnwidth]{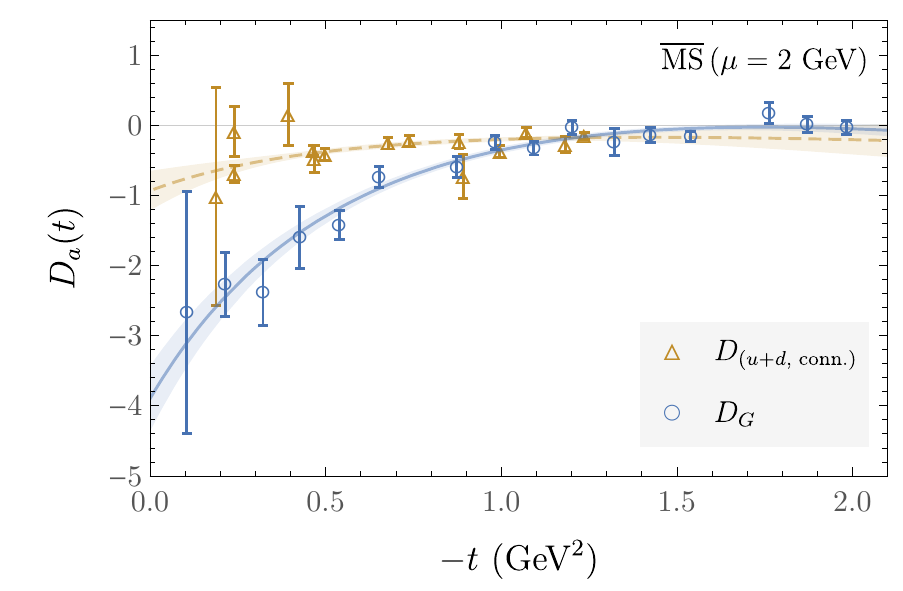}
\caption{\footnotesize $D_G(t)$ and $D_{u+d}(t)$ GFF for the proton from \cite{Shanahan:2018pib} and \cite{LHPC:2007blg} respectively, with quark masses corresponding to a pion mass of approximately 450 MeV. \label{fig:Dg}}
\end{figure}

Beyond forward-limit matrix elements, lattice QCD has also been used to compute the quark and gluon GFFs of the proton and other hadrons. Such calculations are computationally more demanding than those needed to constrain the forward-limit components, and statistical uncertainties increase with $|t|$. As a result, these studies have not yet achieved the same level of systematic control as the spin and mass decomposition. Nevertheless, the quark contributions to the proton's GFFs (and those of other hadrons such as the pion) have been computed with $|t|\lesssim 1~\text{GeV}^2$~\cite{Bali:2016wqg,Brommel:2005ee,Brommel:2007zz,Yang:2018bft,Alexandrou:2018xnp,Yang:2018nqn,Alexandrou:2017oeh,Alexandrou:2020sml,Alexandrou:2019ali,LHPC:2007blg}. The gluon contributions to the proton's GFFs are far less well-constrained, and almost all calculations to date have been performed with quark masses corresponding to larger-than-physical values of the pion mass~\cite{Shanahan:2018nnv,Shanahan:2018pib,Pefkou:2021fni,Detmold:2017oqb}. Nevertheless, the gluon GFFs with $|t|\lesssim 2~\text{GeV}^2$ were computed for a range of hadrons in \cite{Pefkou:2021fni}, allowing qualitative comparisons of their $t$-dependence as illustrated in Fig.~\ref{fig:AgGFFLattice}. Of particular recent interest has been the $D(t)$ GFF, which does not have a sum-rule constraint in the foward limit; a comparison between lattice QCD calculations of the quark and gluon contributions is illustrated in Fig.~\ref{fig:Dg}. 

In contrast to local matrix elements, matrix elements defined with light-cone separations, yielding e.g. the $x$-dependence of GPDs, can not be directly computed in Euclidean spacetime, but must be approached by indirect means. 
Significant developments over the last two decades have 
yielded a range of complementary approaches to direct calculations of GPDs themselves in the lattice QCD framework~\cite{Detmold:2005gg,PhysRevLett.110.262002,Chambers:2017dov,Ma:2017pxb,Radyushkin:2017cyf,Constantinou:2020hdm,Detmold:2021uru}. Given the significant technical and computational challenges of these approaches, the first lattice QCD studies of the $x$-dependence of the proton GPDs were achieved only recently in 2020~\cite{Alexandrou:2020zbe,Lin:2020rxa}. Calculations with complete systematic control will require continued efforts over the coming years. 


\section{Experimental results}
\label{experiments}

This section presents a discussion of the DVCS data and the analysis procedure that led to the first extraction of the proton $D$-term form factor $D_q(t)$ from data collected with the CLAS detector at 
Jefferson Lab (JLab). The extraction of $D_q(t)$ of  {$\pi^0$} from Belle data, and other phenomenological results, are also reviewed.

\subsection{DVCS in fixed-target and collider experiments } 


The first measurements of DVCS on unpolarized protons were carried out with the H1~\cite{H1:2001nez} experiment and later with the ZEUS~\cite{ZEUS:2003pwh}~ experiment, both at the HERA collider.  
The first observation of the $\sin(\phi)$-dependence for the $ {\vec e} p \rightarrow e'p'\gamma $ process as signature of the interference of the DVCS and Bethe-Heitler amplitudes came from the CLAS \cite{CLAS:2001wjj} and HERMES \cite{HERMES:2001bob} detectors.


These initial results triggered the development of a worldwide dedicated experimental program to measure the DVCS process with  high precision and in a large kinematic range with HERMES at HERA, Hall A and CLAS at JLab, and COMPASS at CERN.
A review of the early DVCS experiments can be found in \cite{dHose:2016mda}.

\subsection{First extraction of
the proton GFF \boldmath $D_q(t)$}
\label{Subsec:D-term-at-JLab}

In this section, the data and procedure used in~\cite{Burkert:2018bqq} to obtain the first determination of the quark contribution to the $D$-term of the proton are described. This work is based on two main pieces of experimental 
information from the CLAS detector at JLab~\cite{CLAS:2003umf}, namely the beam-spin 
asymmetry (BSA) measured with spin-polarized electron beams, and the unpolarized cross section for DVCS 
on the proton. 

The polarization asymmetries and differential cross sections have been used to extract the imaginary and real parts of the CFF ${\cal H}$ respectively. Using the 
dispersion relation technique to determine the subtraction term $\mathcal C_{\mathcal H}(t)$, as discussed in section \ref{DVCS}, requires the full integral over $0 \leq \xi \leq 1$ at fixed $t$ to be evaluated. As this process requires an extrapolation to both $\xi=0$ and to $\xi=1$ that are unreachable in experiments, a parameterization of the $\xi$-dependence of Im$\mathcal{H}$ close to these limits  has been incorporated to fit the data. 

In the first step, fits of the BSA~\cite{CLAS:2007clm} and of the unpolarized differential cross-sections~\cite{CLAS:2015uuo} for DVCS were performed to estimate Im${\mathcal H(\xi,t)}$ and Re${\mathcal H(\xi,t)}$ at fixed kinematics in $\xi$ and $t$ in the ranges covered by the data. The BSA is defined as 
\begin{eqnarray}
A_{LU}(\xi, t) = \frac{N^+(\xi, t) - N^-(\xi, t)}{N^+(\xi, t) + N^-(\xi, t) } ,
\end{eqnarray}
where $N^+$ and $N^-$ refer to the  measured event rates at electron helicity $+1$ and $-1$, respectively.

\begin{figure}[t!] 
\includegraphics[width=0.9\linewidth]{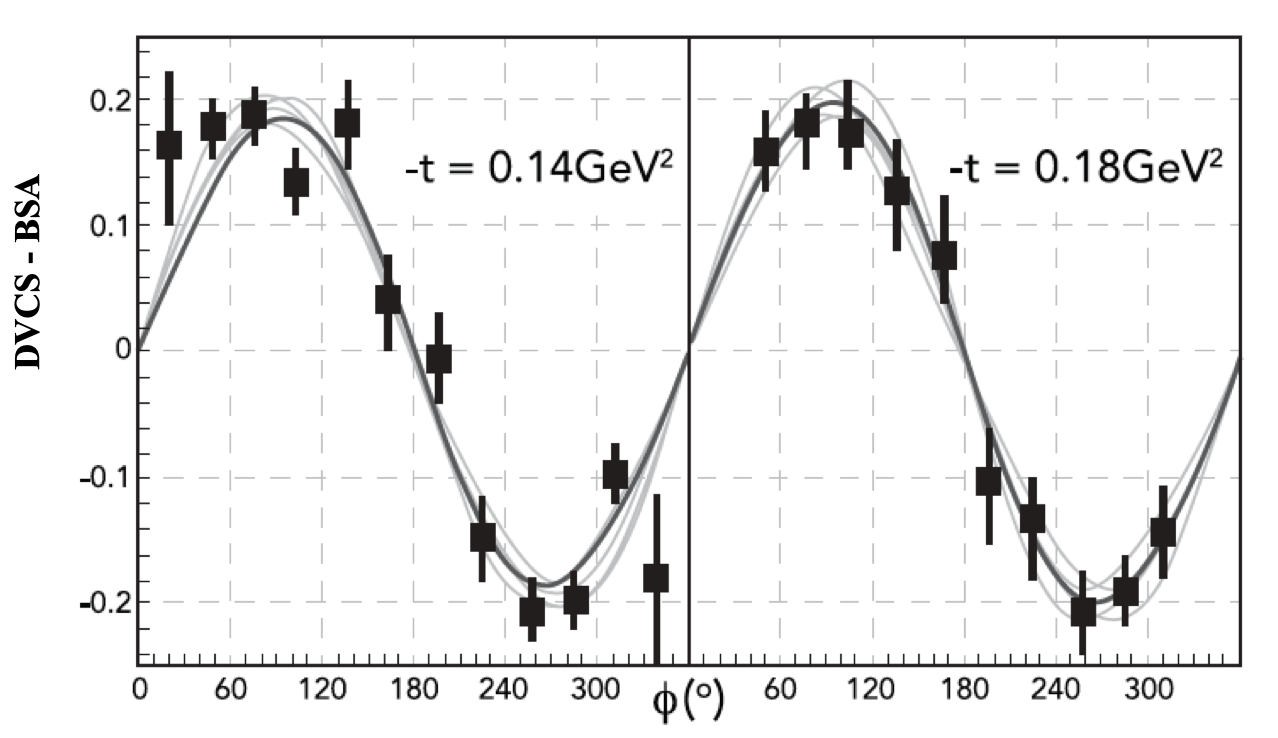}
\includegraphics[width=0.9\linewidth]{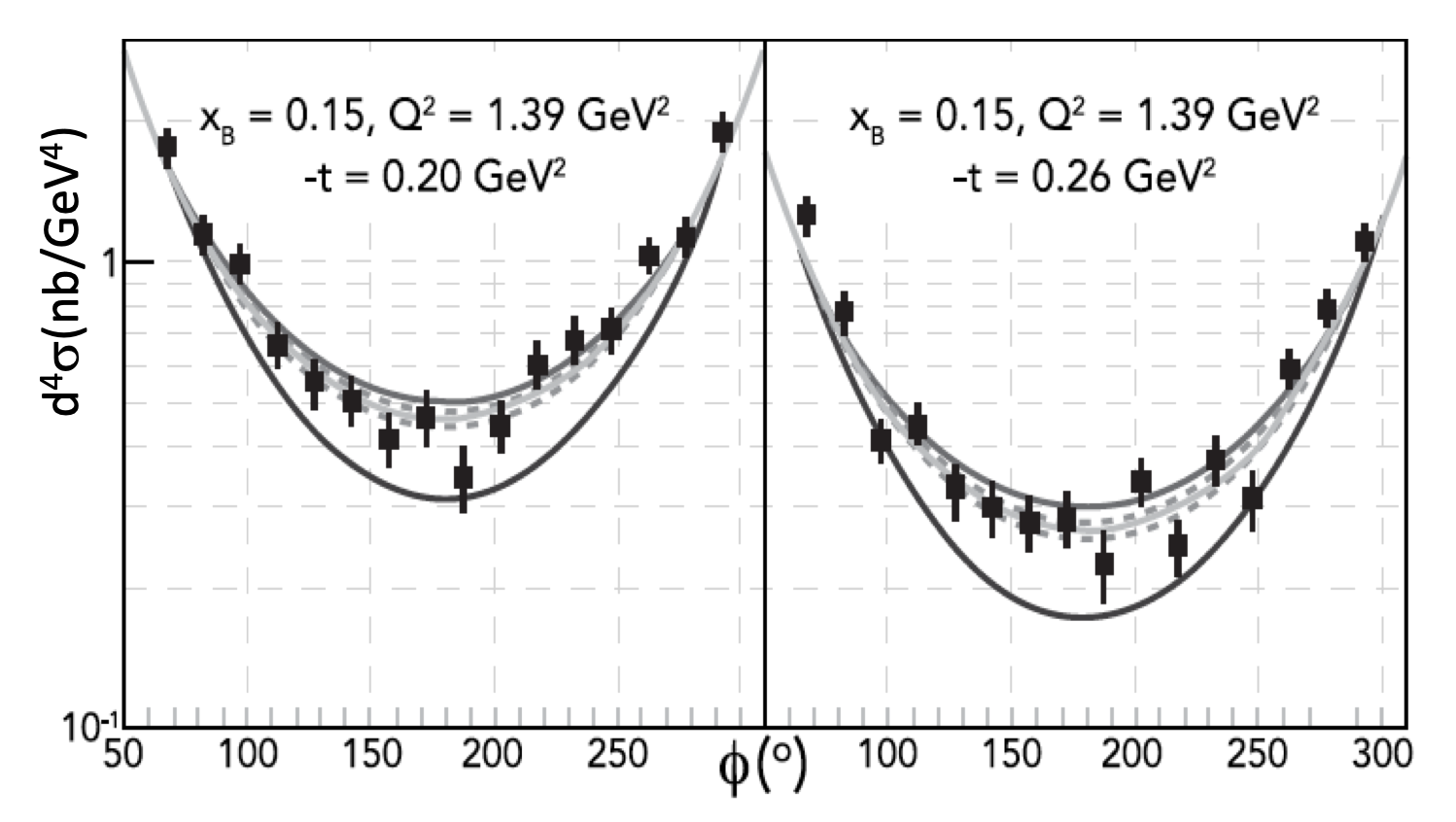}

\vspace{-2mm}

\caption{Top: The expected $\sin{\phi}$ dependence   is fit to the data. The thick solid lines 
are the global fits using the parameterization  
according to~\eqref{km-para}. The bunch of thin solid lines represent 
local fits. The spread of the lines represent estimates of the systematic uncertainties. Bottom: The unpolarized cross section at fixed $\xi$ and $Q^2$ for different $t$ values. The azimuthal $\phi$ angle dependence of the cross section is fit to the experimental data. The thin dark solid line is the global fit. The upper thin gray lines represent fits at the given kinematics with the dashed lines showing the systematic uncertainties. The lower thick black lines show the Bethe-Heitler contributions. The graphics is adapted by permission from~\cite{Burkert:2018bqq}. Note the logarithmic vertical scale.}
\label{BSA-DVCS-BH}
\end{figure}

The experimentally-measured BSA in $\vec{e} p \to e p \gamma$ contains not only the DVCS term, with the photon generated at the proton vertex, but also the Bethe-Heitler term with the photon generated at the incoming or scattered electron, respectively (see Fig.~\ref{DVCS-BH}). Both have the same final state and thus interfere. They generate a $\sin\phi$-dependent interference contribution as seen in Fig.~\ref{BSA-DVCS-BH}. The DVCS term is dominated by the CFF Im$\mathcal{H}$ and the Bethe-Heitler term is real and is given by the elastic electromagnetic FFs.  

It is important to note that this analysis does not rely on extracted cross sections but on asymmetries of event rates in specific bins. This is an essential 
advantage 
as it avoids accounting for systematic uncertainties that must be included in the cross section extraction. The uncertainties in $A_{LU}(\xi,t)$ are dominated by statistics rather than systematic uncertainties, which determines the local values of Im$\mathcal{H}$ very precisely as can be seen in the top panel of Fig. \ref{BSA-DVCS-BH}, which shows the BSA and the differential cross sections for selected kinematic bins. 
  
In the second step, the 
Im$\mathcal{H}(\xi,t)$ are fit with the functional form used in global fits~\cite{Muller:2013jur,Kumericki:2016ehc} with the parameters fit to the local CLAS data. The imaginary part is written as:
\begin{equation}
    \begin{aligned}
{\textrm{Im}}\mathcal{H}(\xi,t) =  \frac{\mathcal N}{1+\xi}\frac{\left(\frac{2\xi}{1+\xi}\right)^{-\alpha(t)}\left(\frac{1-\xi}{1+\xi}\right)^b}{ \left(1-\frac{1-\xi}{1+\xi}\frac{t}{M^2}\right)^{p}}, \label{km-para} 
\end{aligned}
\end{equation}
where $\mathcal N$ is a free normalization constant, $\alpha(t)$ is fixed from small-$x$ Regge phenomenology as $\alpha(t)=0.43+0.85\,t~\text{GeV}^{-2}$,
$b$ is a free parameter controlling the large-$x$ behavior, $p$ is fixed to 1 for the valence quarks, and $M$ is a free parameter controlling the $t$-dependence.

\begin{figure}[ht!]
\vspace{-0.4cm}

\includegraphics[width=0.95\linewidth]{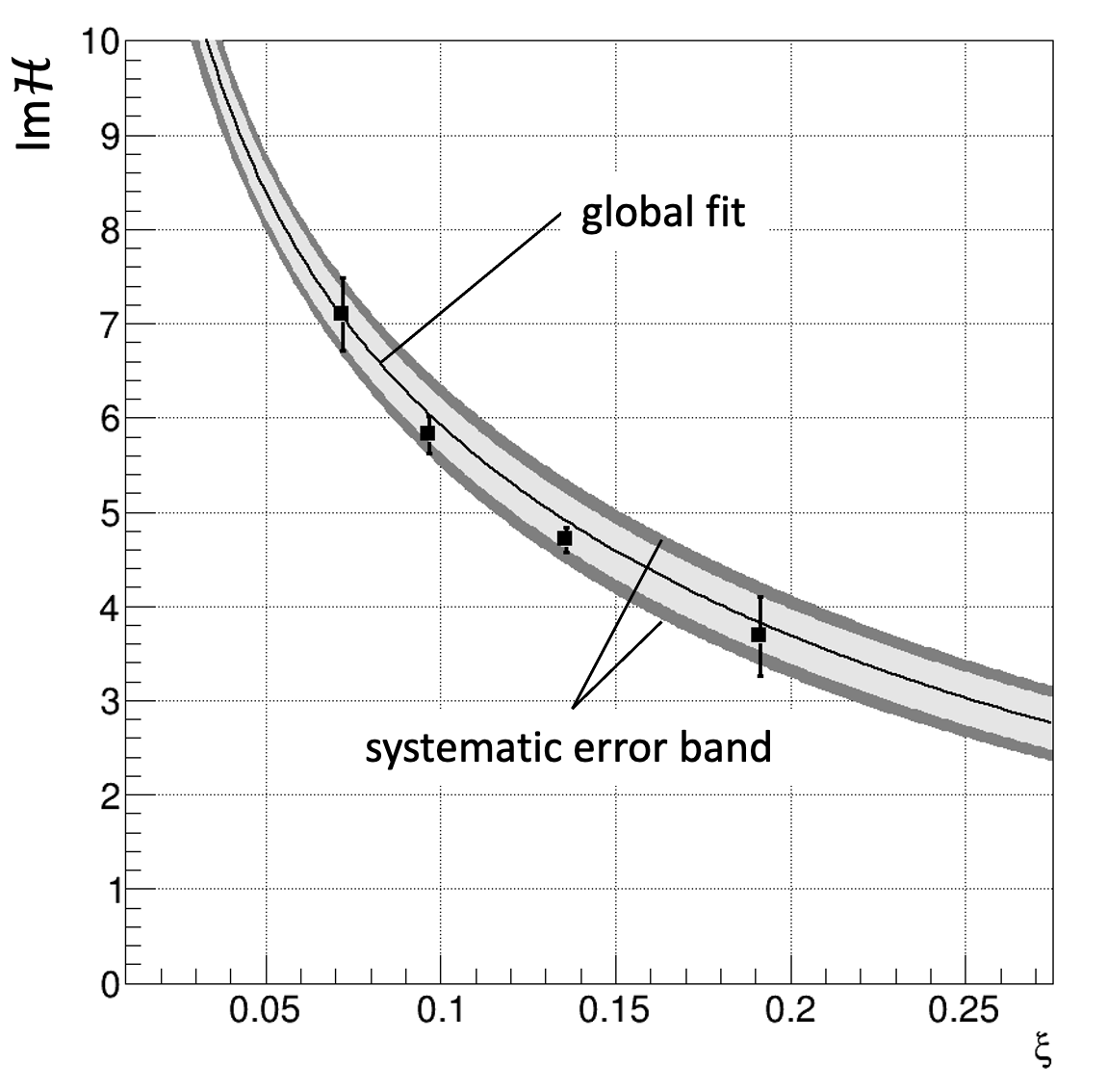}


\includegraphics[width=0.95\linewidth]{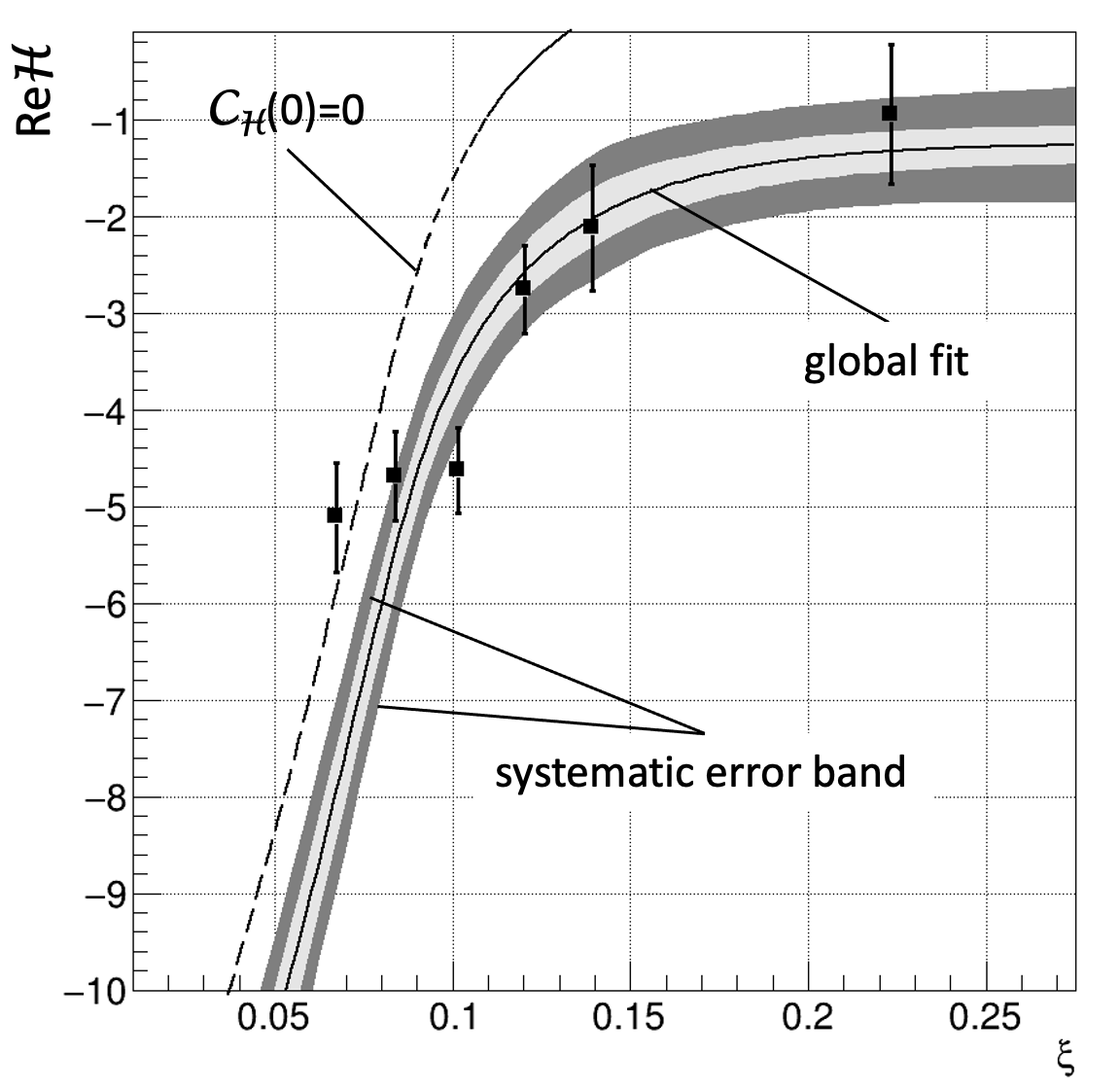}

\vspace{-3mm}

\caption{\footnotesize Top: The Im${\cal H}$ data points are plotted as function of $\xi$ from local fits to the $A_{LU}$ data~\cite{CLAS:2007clm} for 
$-t=0.13$-$0.15$ GeV$^2$. The central solid line is the global fit constrained by the data points. The light gray error band
is due to the uncertainty of the other CFFs. The outer dark-gray band shows the total systematic uncertainty to the imaginary part of the fit.  Bottom: Re$\cal{H}$ data as extracted from unpolarized cross section data~\cite{CLAS:2015uuo}. 
 The central solid curve shows the result of the global fit with the dispersion relation applied and 
the fit parameters of the multipolar form for $\mathcal C_{\mathcal H}(t)$. The other lines/bands describe the same contribution as for Im${\cal H}$ propagated with the dispersion relation.  
The dashed line separated from the error bands shows the real part of the amplitude computed from the imaginary part using the dispersion relation and setting $\mathcal C_{\mathcal H}(0)$ to zero. The difference of dashed line and the solid line shows the effect of the subtraction term. Note that all markers in Re${\cal H}$ contribute to the precision of a single $-t$ value in $\mathcal C_{\mathcal H}(t) $, resulting in a small fit uncertainty.}
\label{CFF}
\end{figure}

The real and imaginary part are fit together including the subtraction term in the dispersion relation~\eqref{DR}.  
Fig.~\ref{CFF} compares the fits at fixed kinematics (local fits) with the global fit 
for one of the $t$ values. The global and local fits show good agreement in $\xi$ and $t$ kinematics where they overlap.    

\begin{figure}[ht]
\includegraphics[width=0.97\columnwidth]{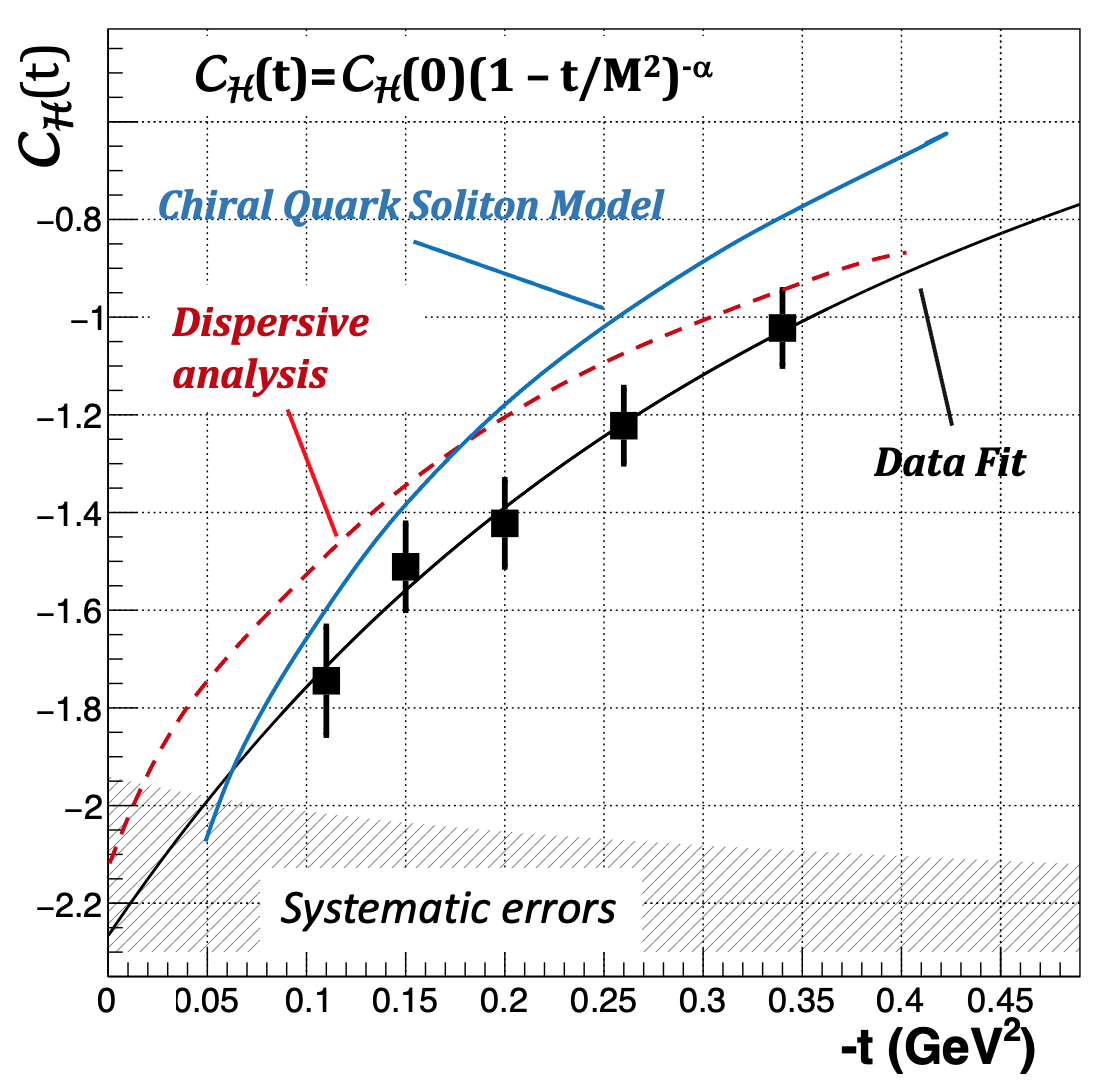}
\caption{\footnotesize The subtraction term $\mathcal C_{\mathcal H}(t)$ as determined from the dispersion relation in the global fit (markers), adapted from~\cite{Burkert:2018bqq}. 
The uncertainties represent results of the fit errors. The hatched area at the bottom represents the estimated 
systematic uncertainties as described in Fig.~\ref{CFF} for one of the bins in $-t$. The dashed and solid-blue curves show the dispersive calculation~\cite{Pasquini:2014vua} and the chiral quark-soliton model predictions~\cite{Goeke:2007fp}, respectively.}
\label{Dt}
\end{figure} 

In the  
fit, $\mathcal C_{\mathcal H}(t)$ is obtained at fixed  
$t$. The results for the subtraction term and the fit to the multipole form
\begin{eqnarray}
\mathcal C_{\mathcal H}(t) &=& \mathcal C_{\mathcal H}(0) \bigg[1 + \frac{(-t)}{M^2}\bigg]^{-\lambda} \label{multipole} \\  \nonumber
\end{eqnarray} 
are displayed in Fig.~\ref{Dt}, where $\mathcal C_{\mathcal H}(0)$, $\lambda$ and $M^2$ are the fit parameters, with their values found to be: 
\begin{equation}
    \begin{aligned}
\mathcal C_{\mathcal H}(0) &=-2.27\pm 0.16 \pm 0.36,  
\\ 
M^2 &=1.02\pm  0.13 \pm 0.21  {\rm ~GeV^2},  \\ 
 \lambda &=2.76 \pm  0.23 \pm 0.48. 
\end{aligned}
\label{parameters}
\end{equation} 
 \noindent The first error is the fit uncertainty, and the second error is due to the systematic uncertainties.
Adding the fit errors for $\mathcal C_{\mathcal H}(0)$ and the systematic errors in quadrature $\sigma_{\mathcal C_{\mathcal H}(0) } = \sqrt{0.16^2 + 0.36^2} \approx  0.39$, 
the significance $S$ of the knowledge of the subtraction term is: 
\begin{eqnarray}
S = \frac{\mathcal C_{\mathcal H}(0) }{\sigma_{\mathcal C_{\mathcal H}(0)}} \approx 5.8.
\end{eqnarray}
More flexible analyses based on unconstrained artificial neural network techniques~\cite{Kumericki:2019ddg,Dutrieux:2021nlz} find however that a more conservative extraction of the subtraction constant from the currently available experimental data remains compatible with zero within large uncertainties.

\begin{figure}[th!] 
\includegraphics[width=1.0\linewidth]{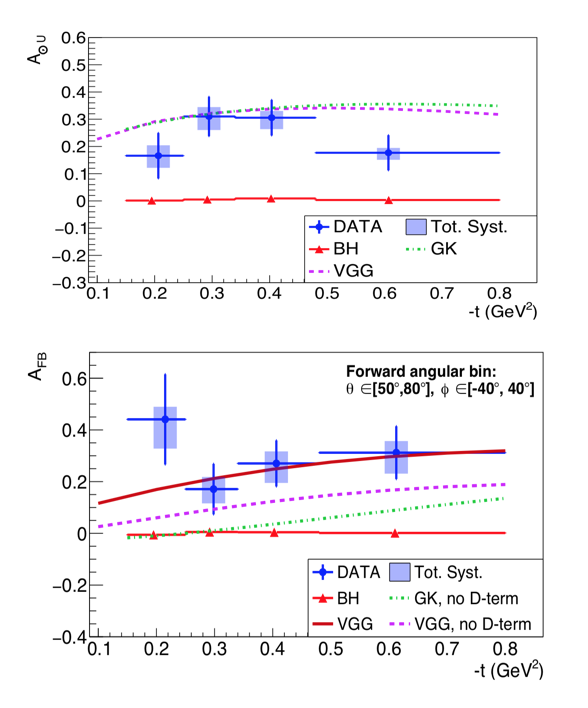}
\caption{The TCS polarized BSA (top) and the TCS $A_{FB}$ (bottom) for an average  1.8~GeV mass of the time-like photon $M_{e^+e^-}$. A value for $A_{LU}$ of (20-25)\% is consistent with what is measured in DVCS and projects out Im${\cal H}$. The FBA projects out Re${\cal H}$ that relates directly to the protons $D_q(t)$-term. Graphics adapted from ~\cite{CLAS:2021lky}. 
The data require the presence of the $D$-term as seen in the difference of the dashed magenta line and the solid red line. At the kinematics of the data in Fig.~\ref{Dt},
about half of the asymmetry may be due to the $D$-term 
when comparing calculations without and with the 
$D$-term~\cite{Vanderhaeghen:1999xj,Pasquini:2014vua}.}
\label{TCS-BSA-FBA}
\end{figure}
In the analysis of~\cite{Burkert:2018bqq}, the term $d_3^q(t)$ and other higher-order terms have been omitted in the expansion~\eqref{Gegenbauer} to extract the GFF $D_q(t)$. The estimated effect is included in the systematic error analysis. It is also assumed that $u$ and $d$ quarks have the same first moments $d_1^u \approx d_1^d \approx {d_1^{u+d} /2}$, an assumption justified in the large-$N_c$ limit~\cite{Goeke:2001tz}. Under these approximations, it follows from~\eqref{Dtermint} that
\begin{eqnarray}
\mathcal C_{\mathcal H}(t)\approx\frac{10}{9}\,d_1^{u+d}(t) = \frac{25}{18}\,D_{u+d}(t).    
\label{F7}
\end{eqnarray}
The truncation in~\eqref{Gegenbauer} leads to a systematic uncertainty of a priori unknown magnitude. For $Q^2\to\infty$, the higher order terms $d^q_3, d^q_5,\cdots$ vanish. But at the $Q^2$ that can be reached in the current experiments, they are not necessarily negligible.    
The results of the chiral quark-soliton model, which predicts values of $d^{u+d}_1$ close to findings
in the experimental analysis~\cite{Goeke:2007fp}, can been used to estimate the contribution of the $d_3^q$ term. At the kinematics relevant for this analysis a ratio $d^{u+d}_3/d^{u+d}_1 \approx 0.3$ was found~\cite{Kivel:2000fg}. A systematic uncertainty of $\delta(d^{u+d}_1)/d^{u+d}_1 =\pm 0.30$ has therefore been included into the results of~\cite{Burkert:2018bqq} for $d_1^{u+d}(t)$.

One may ask if the first two terms in the Gegenbauer polynomial expansion $d_1^q(t)$ and $d_3^q(t)$ could be separated in some way to reduce the systematics. This has been studied in~\cite{Dutrieux:2021nlz} by including the $Q^2$-evolution into the phenomenological analysis. It was found that, assuming the same $t$-dependence, the two terms cannot currently be separated given the limited range in $Q^2$ covered by the data. In the future one may expect Lattice QCD to be able to provide a model-independent evaluation of this higher-order contribution.

To conclude this section, the determination of $\mathcal C_{\mathcal H}(t)$ suggests that the quark contribution $\sum_qD_q(t)$ to the proton's GFF $D(t)$ is non-zero and large. These results have been supported in a recent paper on the first measurement of TCS~\cite{CLAS:2021lky} as shown in Fig.~\ref{TCS-BSA-FBA}, where the contribution of the $D$-term to the forward-backward asymmetry is seen to be 
significant. Moreover, predictions in the 
chiral quark-soliton model \cite{Goeke:2007fp} and from dispersive analysis~\cite{Pasquini:2014vua} shown in Fig.~\ref{Dt} are consistent with the results discussed here within the systematic uncertainties.

\subsection{Other measurements and phenomenological studies}

The first extraction of the $\pi^0$ GFFs in the time-like region based on the process $\gamma \gamma^* \to \pi^0 \pi^0$, depicted in Fig.~\ref{Jpsi-gamma-gamma-pi-pi}, which was measured in the Belle experiment in $e^+e^-$ collisions \cite{Belle:2015oin}, was obtained in \cite{Kumano:2017lhr}.
For the quark contribution to the $\pi^0$ $D$-term 
the value $\sum_qD_q(0)\approx - 0.75$ was reported, 
but systematic uncertainties have not been estimated.
It has recently been observed in~\cite{Lorce:2022tiq}
that kinematical corrections may significantly impact 
the extraction of generalized distribution amplitudes
from experimental data and should be taken into 
account in future analyses.

{
Based on data from experiments at JLab on the energy-dependence of $J/\Psi$ production at threshold~\cite{GlueX:2019mkq,Duran:2022xag}, phenomenological information on the gluon $D_G(t)$ form factor of the proton was extracted
\cite{Kharzeev:2021qkd,Kou:2021qdc,Wang:2022ndz} and estimates were obtained for the  gluon contributions to the proton mean square mass radius $\int\ud^3r\,r^2\mathcal T^{00}(\vec r)/M_N$ and the mean square scalar radius $\int\ud^3r\,r^2g_{\mu\nu}\mathcal T^{\mu\nu}(\vec r)/M_N$. The most recent data on this process were reported in \cite{GlueX:2023pev}.}
(For remarks on the theoretical status of this process
see Sec.~\ref{meson production}.) 
A similar study for the lighter
$\phi$-meson was presented in \cite{Hatta:2021can}.

\subsection{Future experimental developments to access GFFs}

As discussed in section III, measurements of DVCS have so far been most effective in obtaining information related to GPDs. However, there are different experimental processes that may be employed to provide additional, or independent, information on the GPDs and GFFs.  

Implementation of a high-duty-cycle positron source, both polarized and unpolarized~\cite{PEPPo:2016saj}, at JLab would significantly enhance its capabilities in the extraction of the CFF  Re$\mathcal{H}(\xi,t)$ and thus of the gravitational form factor $D_q(t)$ and of the mechanical properties of the proton.  

The time-like Compton scattering process will be measured in parallel to the DVCS process employing the large acceptance detector systems such as CLAS12~\cite{Burkert:2020akg}. The TCS event rate is much reduced compared to DVCS and requires higher luminosity for similar sensitivity to $\mathcal{H}$. 
In experiments employing large acceptance detector 
systems, both the DVCS and TCS processes are measured 
simultaneously, in quasi-real photo-production at very small $Q^2 \to 0$, and in real photo-production, where
the external production target acts as a radiator of 
real photons that undergo TCS further downstream 
in the same target cell. 

The double DVCS process enables access to GPDs in their full kinematic dependencies on $x, \xi, t$, see Sec.~\ref{sec:EMT-FF}. At the same time it is reduced in rate by orders of magnitude compared to 
DVCS~\cite{Kopeliovich:2010xm}  
requiring  
higher luminosity than is currently achievable. 
Nevertheless, special equipment that would 
comply with such requirements has been proposed~\cite{Chen:2014psa}. Such measurements are currently planned at JLab in Hall A and  
Hall B.

Finally, an energy-doubling of the existing electron accelerator at JLab is currently under consideration~\cite{Arrington:2021alx}. This upgrade would   extend the DVCS program to higher $Q^2$ and lower $x_B$ and better link the DVCS measurements at the current 12 GeV operation to the kinematic reach that will be available at the Electron-Ion Collider, a flagship future facility in preparation at the Brookhaven National Laboratory (discussed further below). It will also more fully open the charm sector to access the gluon GFFs. 

\section{Interpretation}
\label{interpretation}

In section~\ref{sec-II} various properties of the GFFs 
have been discussed at zero momentum transfer. Much of the
recent interest in GFFs comes from the fact that they 
contain information on the spatial 
distributions of energy, angular momentum, and internal forces 
that can be accessed at non-zero momentum transfer $t$, via 
an appealing interpretation which is reviewed here.

\subsection{\boldmath The static EMT}
\label{Sec:static-EMT}

The 3D interpretation 
{\cite{Polyakov:2002yz} in Eq.}~\eqref{Eq:static-EMT} 
of the information encoded by GFFs provides 
analogies to intuitive concepts such as pressure. 
A 2D interpretation can also be carried out in other 
frames~\cite{Lorce:2018egm,Freese:2021czn,Freese:2021mzg} 
with Abel transformations allowing one to relate 2D and 
3D interpretations~\cite{Panteleeva:2021iip}.

Considering 2D EMT distributions for a nucleon state 
boosted to the infinite-momentum frame has the advantage 
that in this case the {nucleon can be perfectly localized around the} transverse center of momentum \cite{Burkardt:2000za}. 
In other frames or in 3D, an exact probabilistic parton 
density interpretation does not hold in general.
The reservations are analogous to those in the case of, 
e.g., the interpretation of the electric FF in terms of 
a 3D electrostatic charge distribution (and the definition 
of electric mean square charge radius which, despite all
caveats, remains a popular concept, giving an idea of the
proton's size).
The 3D EMT description is nevertheless mathematically 
rigorous \cite{Polyakov:2018zvc} and can be interpreted 
in terms of quasi-probabilistic distributions from a 
phase-space point of view \cite{Lorce:2018egm,Lorce:2020onh}.  
A strict probabilistic interpretation is, however, justified 
for heavy nuclei and for the nucleon in the large-$N_c$ 
limit, where recoil effects can be safely neglected  
\cite{Polyakov:2002yz,Goeke:2007fp,Polyakov:2018zvc,Lorce:2022cle}.

The meaning of the different components of the static EMT is 
intuitively clear, with $\mathcal T^{00}(\vec{r})$ denoting 
the energy distribution and $\mathcal T^{0k}(\vec{r})$ 
representing the spatial distribution of  momentum.  
In the following sections the focus is on 
$\mathcal T^{ij}(\vec{r})$ which are perhaps the most 
interesting components of the static EMT, thanks to 
their relation to the stress tensor and the $D$-term.

\subsection{\boldmath The stress tensor and the $D$-term}
\label{Sec:stress-tensor}

The key to the mechanical properties of the proton is the 
symmetric stress tensor $\mathcal T^{ij}(\vec{r})$ given 
by \cite{Polyakov:2002yz} 
\begin{equation}
    \label{Eq:Tij-p-s}
    \mathcal T^{ij}(\vec{r}) = 
    \biggl(\frac{r^ir^j}{r^2}-\frac13\,\delta^{ij}\biggr)\,s(r)
    + \delta^{ij}\,p(r)\,
\end{equation}
with $s(r)$ known as the shear force (or anisotropic stress) 
and $p(r)$ as the pressure with $r=|\vec{r}|$. 
Both are connected by the differential equation 
$\frac23\,\frac{\ud}{\ud r}s(r)+\frac2r\,s(r)+\frac{\ud}{\ud r}p(r)=0$ 
and $p(r)$ obeys $\int_0^\infty \ud r\,r^2p(r)=0$ 
\cite{von-Laue:1911}, a necessary but not sufficient 
condition for stability. These relations originate from 
the EMT conservation expressed by 
$\nabla^i\mathcal T^{ij}(\vec{r}) = 0$ for the static EMT. 
The total $D$-term $D(0)$ can be expressed in terms of 
$p(r)$ and $s(r)$ in two equivalent ways,
\begin{eqnarray} 
    D(0) 
    = - \frac{4}{15}\,M_N \!\!\int{\ud^3r}\, r^2 s(r) 
    = M_N\!\!\int{\ud^3r}\,r^2 p(r) \label{Eq:D-term} 
    \,. \;\;\;\;\;
\end{eqnarray} 
The form of the stress tensor~\eqref{Eq:Tij-p-s} is valid 
for spin-0 and spin-$\frac12$ hadrons; for higher spins
see~\cite{Cosyn:2019aio,Polyakov:2019lbq,Cotogno:2019vjb,Kim:2020lrs,Ji:2021mfb}.

If the GFF $D(t)$ is known, then $s(r)$ and $p(r)$ 
are obtained as follows \cite{Polyakov:2018zvc}
\begin{eqnarray} 
    s(r) &=& -\frac{1}{4M_N}\,r\,\frac{\ud}{\ud r}
    \frac{1}{r}\frac{\ud}{\ud r}\widetilde{D}(r),\label{sr-II} \\
    p(r) &=& \frac{1}{6M_N}\frac{1}{r^2}\frac{\ud}{\ud r}
    r^2\frac{\ud}{\ud r}\widetilde{D}(r),\label{pr-II}
\end{eqnarray} 
where $\widetilde{D}(r)= \int\frac{\ud^3\Delta}{(2\pi)^3}\,e^{-i{\vec\Delta\cdot\vec r}} D(-{\vec \Delta}^2)$. 
If the separate $D_q(t)$ and $D_G(t)$ GFFs are known,
``partial'' quark and gluon shear forces $s_q(r)$ and 
$s_G(r)$ can be defined in analogy to (\ref{sr-II}). 
In order to define ``partial'' quark and gluon pressures, 
in addition to $D_q(t)$ and $D_G(t)$ knowledge of 
$\bar{C}_q(t)=-\bar{C}_G(t)$ is required. 
The latter are responsible for ``reshuffling'' forces
between the gluon and quark subsystems inside the proton
\cite{Lorce:2017xzd,Polyakov:2018exb} and are difficult to access 
experimentally. 
$\bar{C}_q(t)$ was studied in the bag model \cite{Ji:1997gm}, chiral quark-soliton model \cite{Goeke:2007fp}, instanton vacuum model \cite{Polyakov:2018exb} and lattice QCD \cite{Liu:2021gco}. Estimates guided by renormalization group methods \cite{Hatta:2018sqd,Tanaka:2018nae,Ahmed:2022adh} yield $\bar{C}_q(0)=-0.163(3)$ at $\mu = 2\,{\rm GeV}$ in $\overline{\text{MS}}$ scheme \cite{Tanaka:2022wzy}.

\subsection{\boldmath Normal forces and the sign of the $D$-term}
\label{Sec:D-term-sign}

The stress tensor $\mathcal T^{ij}(\vec{r})$ can be 
diagonalized, with one eigenvalue given by the normal 
force per unit area $p_n(r)=\frac23\,s(r)+p(r)$ with 
the pertinent eigenvector $\vec{e}_r$. The other two
eigenvalues are degenerate (for spin-0 and spin-$\frac12$) 
and are known as tangential forces per unit area, 
$p_t(r)=-\,\frac13\,s(r)+p(r)$, with eigenvectors 
which can be chosen to be unit vectors in the 
$\vartheta$- and $\varphi$-directions in 
spherical coordinates \cite{Polyakov:2018zvc}.

The normal force appears when considering the force 
$F^i=\mathcal T^{ij}\ud S^j=p_n(r)\,\ud S\,e_r^i=[\frac23\,s(r)+p(r)]\,\ud S\,e_r^i$ 
acting on a radial area element $\ud S^j = \ud S\,e_r^j$,
where $e_r^j=r^j/r$.
General mechanical stability arguments require this force 
to be directed towards the outside, or else the system would 
implode. This implies that the normal force per unit area 
must be positive 
\begin{equation}
    p_n(r) = \frac23\,s(r)+p(r)>0 
    \label{Eq:normal-force-positivity}\,.
\end{equation}
As an immediate consequence of 
(\ref{Eq:normal-force-positivity}) 
one concludes by means of Eq.~\eqref{Eq:D-term} 
that \cite{Perevalova:2016dln}
\be\label{Eq:D-sign}
      D(0) < 0\,.
\ee
For hadronic systems like protons, hyperons, mesons or 
nuclei for which the $D$-term has been computed 
(in models, chiral perturbation theory, lattice QCD or 
by dispersive techniques, see Sec.~\ref{Sec-5:theory}) 
or inferred from experiment (in the case of the proton 
and $\pi^0$, see Sec.~\ref{experiments}) it has always 
been found to be negative in agreement with 
(\ref{Eq:D-sign}).

The above definitions and conclusions are more than 
just a fruitful analogy to mechanical systems. At this 
point it is instructive to recall how one calculates 
the radii of neutron stars, which are amenable to an
unambiguous 3D interpretation. 
In these macroscopic hadronic systems, general relativity
effects cannot be neglected and are incorporated in the
Tolman-Oppenheimer-Volkoff equation, which is solved by 
adopting a model for the nuclear matter equation of state.
The solution yields (in our notation) $p_n(r)$ inside the
neutron star as function of the distance $r$ from the center.
The obtained solution is positive in the center and decreases 
monotonically until it drops to zero at some $r=R_\ast$, 
and would be negative for $r>R_\ast$ corresponding to a
mechanical instability. 
This is avoided and a stable solution is obtained by 
defining $r=R_\ast$ to be the radius of the neutron star, 
see for instance \cite{Prakash:2000jr}. 
Thus, the point where the normal force per unit area drops 
to zero coincides with the ``edge'' of the system. 

The proton has of course no sharp ``edge'', being 
surrounded by a ``pion cloud'' due to which the normal 
force does not drop literally to zero but exhibits a 
Yukawa-type suppression at large $r$ proportional to 
$\frac{1}{r^6}\,e^{-2m_\pi r}$ \cite{Goeke:2007fp}.
In the less realistic but very instructive 
bag model, there is an ``edge'' at the bag boundary, 
where $p_n(r)$ drops to zero \cite{Neubelt:2019sou}. 
In contrast to the neutron star one does not determine the 
``edge'' of the bag model in this way. Rather the normal 
force drops ``automatically'' to zero at the bag radius 
which reflects the fact that from the very beginning 
the bag model was constructed as a simple but mechanically
stable model of hadrons \cite{Chodos:1974je}.

\subsection{\boldmath The mechanical radius of the proton and neutron}
\label{Sec:r-mech}

The ``size'' of the proton is commonly defined through 
the electric charge distribution which is indeed a useful 
concept, though only for charged hadrons. 
For an electrically neutral hadron like the neutron, 
the particle size cannot be inferred in this way.
In that case, one may still define an electric
mean square charge radius $r_{\rm ch}^2 = 6\,G'_E(0)$ 
in terms of the derivative of the electric FF $G_E(t)$
at $t=0$. But for the neutron $r_{\rm ch}^2<0$ which
gives insights about the distribution of electric 
charge inside neutron, but not about its size. 
This is ultimately due to the neutron's charge 
distribution not being positive definite.

The positive-definite normal force per unit area, 
~(\ref{Eq:normal-force-positivity}), is an ideal quantity to define
the size of the nucleon. One can define the {\it mechanical radius} 
as \cite{Polyakov:2018guq,Polyakov:2018zvc}
\begin{equation}
\label{Eq:mech-r}
r_{\rm mech}^2 
= \frac{\int \ud^3r\,r^2\,p_n(r)}
       {\int \ud^3r\,p_n(r)} 
= \frac{6 D(0)}{\int_{-\infty}^0\ud t\,D(t)}\,.
\end{equation}
Interestingly, this is an ``anti-derivative'' of a GFF 
as compared to the electric mean square charge radius defined 
in terms of the derivative of the electric FF at $t=0$.
With this definition, the proton and neutron have the same
radius (modulo isospin violating effects). Notice also
that the (isovector) electric mean square 
charge radius diverges in the chiral limit and is therefore inadequate 
to define the proton size in that case, 
while the mechanical radius in~(\ref{Eq:mech-r}) 
remains finite in the chiral limit \cite{Polyakov:2018zvc}. 
The mechanical radius of the proton is predicted to be somewhat smaller 
than its charge radius in soliton models \cite{Goeke:2007fp,Cebulla:2007ei}. 
The charge and mechanical radii become equal in 
the non-relativistic limit 
which was derived in the bag model \cite{Neubelt:2019sou,Lorce:2022cle}.

\subsection{First visualization of forces from experiment}

\begin{figure}[bhp!]
\includegraphics[width=1.0\columnwidth]{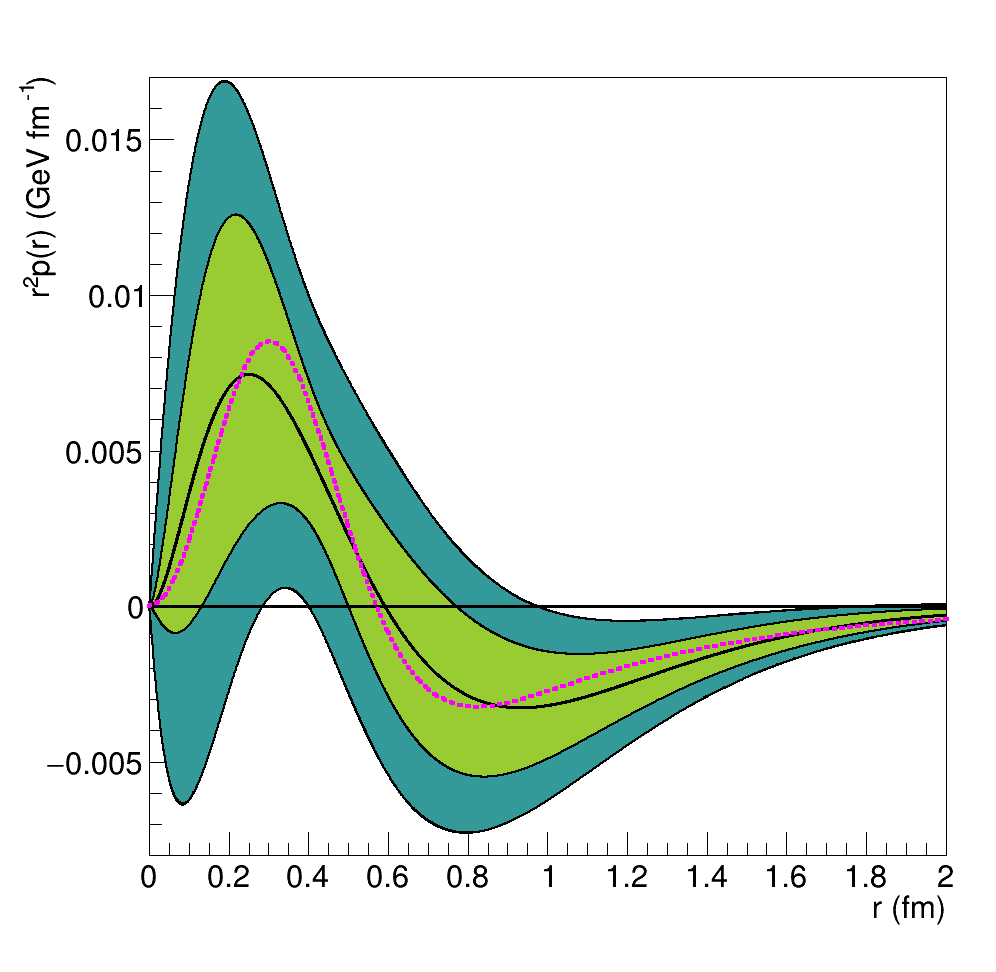} 
\includegraphics[width=1.0\columnwidth]{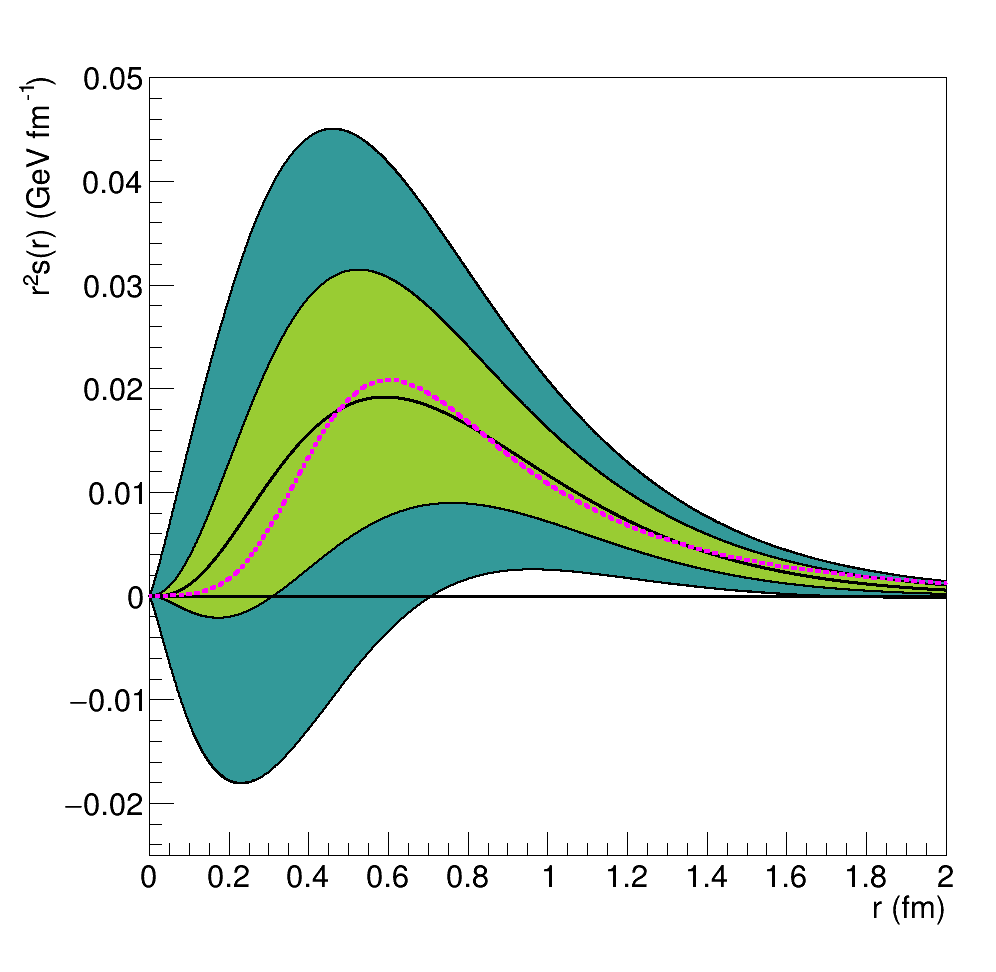}
\caption{\footnotesize The distributions of pressure 
$r^2p_q(r)$ (top) and shear stress $r^2s_q(r)$ (bottom) 
on quarks in the proton 
based on JLab data \cite{Burkert:2018bqq,Burkert:2021ith}.
The central solid lines show the best 
fit.  The outer shaded areas mark the uncertainties when 
only data prior to the CLAS data are included. 
The inner shaded areas represent the uncertainties 
when the CLAS data are used. The widths of the bands 
are dominated by systematic uncertainties [which include extrapolation in unmeasured $\xi$-region when 
evaluating~(\ref{DR}) and the neglect of higher-order terms 
in the Gegenbauer expansion described in~\eqref{F7}]. 
The dotted magenta curves represent the model predictions
of~\cite{Goeke:2007fp}.}
\label{pressure-shear}
\end{figure} 

\begin{figure*}[thbp!]
\includegraphics[width=2\columnwidth]{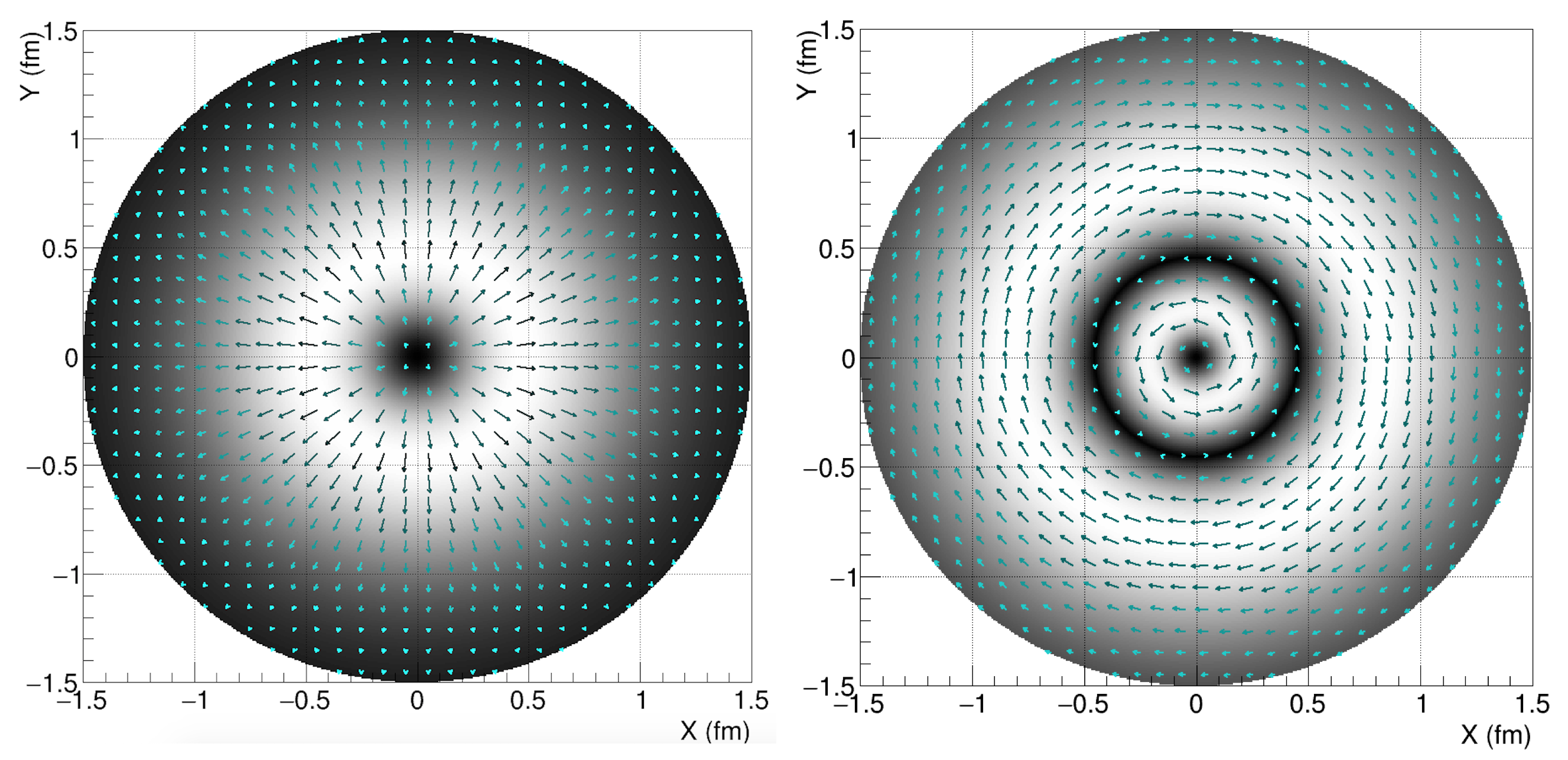}
\caption{\footnotesize 2D display of the quark contribution to the distribution of forces in the proton as a function of the distance from the proton's center~\cite{Burkert:2021ith}. The light gray shading and longer arrows indicate areas of stronger forces, the dark shading and shorter arrows indicate areas of weaker forces. Left panel: Normal forces as a function of distance from the center. The arrows change magnitude and point always radially outwards. Right panel: Tangential forces as a function of distance from the center. The forces change direction and magnitude as indicated by the direction and lengths of the arrows. They change sign near 0.4~fm from the proton center. }
\label{normal-tangential-force}
\end{figure*}

The first visualization of the 
force distributions in the proton was presented in \cite{Burkert:2018bqq} which will be reviewed here.
As detailed in Sec.~\ref{Subsec:D-term-at-JLab}, 
the DVCS data from JLab experiments 
\cite{CLAS:2007clm,CLAS:2015uuo} provided  
information on the observable $\mathcal C_{\mathcal H}(t)$
in~\eqref{DR}, from which, under certain reasonable 
(at present necessary) assumptions, information about 
the quark contribution $D_{u+d}(t)$ of the proton 
was deduced. Based on this information, \eqref{pr-II}  
yields the results for the pressure $p_q(r)$ and the 
shear force $s_q(r)$ of quarks displayed in 
Fig.~\ref{pressure-shear}
(the index $q$ denotes here $u+d$ quark contributions, 
with heavier quarks neglected). In order to obtain 
$p_q(r)$, the additional assumption was made that
$\bar C_q(t)$ can be neglected.

The $r^2p_q(r)$ distribution is positive,  peaks near 
$0.25\,$fm, changes sign near $0.6\,$fm, and reaches 
its minimum value around 1.0$\,$fm. The peak value of
$r^2s_q(r)$ is around $20$~MeV\,fm$^{-1}$, and occurs 
near $0.6\,$fm from the proton's center, where the shear
force, given by $4\pi r^2s_q(r)$, reaches $240$~MeV
fm$^{-1}$ or $38$ kN, an appreciably strong force inside
the tiny proton. It is interesting to observe that these 
results are consistent with predictions from the chiral 
quark-soliton model \cite{Goeke:2007fp}
within the (large) systematic uncertainties in the data.

The quark contribution to the normal and tangential forces, 
$p_{n,q}$ and $p_{t,q}(r)$ as defined in
Sec.~\ref{Sec:D-term-sign},
are displayed in a two-dimensional plot in 
Fig.~\ref{normal-tangential-force}. This figure shows the 
3D distributions inside the proton in a slice going through 
the ``equatorial plane''.
The normal forces are strongest at mid-distances near $0.5\,$fm 
from the proton center and drop towards the center and towards 
the outer periphery. The tangential forces exhibit a node near
$0.40\,$fm from the center.

\subsection{\boldmath The $D$-term and long-range forces}

Among the open questions in theory is the issue of how to define 
the $D$-term in the presence of long-range forces. 
It was shown in a classical model of the proton
\cite{Bialynicki-Birula:1993shm} that $D(t)$ diverges like 
$1/\sqrt{-t}$ for $t\to0$ due to the $\frac1r$-behavior of 
the Coulomb potential  \cite{Varma:2020crx}. This result is 
model-independent and was found also for charged pions 
in chiral perturbation theory \cite{Kubis:1999db}, in 
calculations of quantum corrections to the Reissner-Nordstr\"om 
and Kerr-Newman metrics \cite{Donoghue:2001qc},
and for the electron in QED \cite{Metz:2021lqv}. 

The deeper reason why $D(t)$ diverges for $t\to0$ 
due to QED effects might be ultimately related 
to the presence of a massless physical state (the
photon) which has profound consequences in a theory.
Notice that $D(t)$ is the only GFF which exhibits 
this feature when QED effects are included. There are 
two reasons for this. First, the other proton GFFs 
are constrained at $t=0$, see~\eqref{Tconstraint}
and~\eqref{Lconstraint}, while $D(t)$ is not.
Second, $D(t)$ is the GFF most sensitive to 
forces  in a system~\cite{Hudson:2017xug}.
Notice that $D(t)$ is multiplied by the prefactor 
$(\Delta^\mu\Delta^\nu - g^{\mu\nu}\Delta^2)$ such
that despite the divergence of $D(t)$ due to QED 
effects the matrix element 
$\langle p'|T_a^{\mu\nu}|p\rangle$ is 
well-behaved in the forward limit.

There have been studies of the $D$-term for the H-atom
\cite{Ji:2021mfb,Ji:2022exr}, which defy the interpretation
presented here. This is perhaps not a surprise considering 
the differences between hadronic and atomic bound states.  
Atoms are comparatively large, low-density objects.
Pressure concepts from continuum mechanics might not 
apply to atoms whose stability is well-understood within 
non-relativistic quantum mechanics.
In contrast to this, the proton as a QCD bound state has 
nearly the same mass as an H-atom but a much smaller size
$\sim10^{-15}$m and constitutes a compact high-density system 
(15 orders of magnitude more dense than an atom) 
where continuum mechanics concepts may be applied 
and provide insightful interpretations. 
Another important aspect might be played
by the role of confinement absent for atoms which 
can be easily ionized. Hadrons constitute 
a much different type of bound state in this
respect. More theoretical work is needed to
clarify these issues.

\section{Summary and outlook}
\label{conclusions}

This Colloquium gives an overview of the exciting recent developments along a new avenue of experimental and theoretical studies of the gravitational structure of hadrons, especially the proton. 

The gravitational form factors of the proton rose to prominence after the
works of Xiangdong Ji \cite{Ji:1994av,Ji:1996ek} 
illustrated how they can be used to gain insight
into fundamental questions such as: how much do quarks and gluons
contribute to the mass and the spin of the proton? Soon afterwards, Maxim Polyakov~\cite{Polyakov:2002yz} showed that they also provide information about the spatial distribution of mass and spin, and allow one to study the forces at play in the bound system.
These works triggered many follow-up studies and
investigations which have deepened our understanding
of proton structure.

Through matrix elements of the energy-momentum operator, the gravitational form factors of the proton and other hadrons have been studied in theoretical approaches including a wide range of models and in numerical calculations in the framework of lattice QCD. 
In broad terms, the simplest aspects of the EMT structure of the proton and other hadrons (such as the pion) have been understood from theory for some number of years, and first-principles calculations providing complete and controlled decompositions of the proton's mass and spin, for example, are now available. On the other hand, more complicated aspects of proton and nuclear structure, such as gluon gravitational form factors, the $x$-dependence of generalized parton distributions, and energy-momentum tensor matrix elements in light nuclei, have been computed for the first time in the last several years, as yet without complete systematic control, and significant progress can yet be expected over the next decade. Theory insight into these fundamental aspects of proton and nuclear structure is thus currently in a phase of rapid progress, complementing the improvement of experimental constraints on these quantities and, importantly, providing predictions which inform the target kinematics for future experiments.

The first experimental results, discussed in this colloquium, are based on precise measurements of the deeply virtual Compton scattering process with polarized electron beam, that determined both, the beam-spin asymmetry and the absolute differential cross section of $\vec{e}p \to ep\gamma$. Measurements covered a limited range in the kinematic variables which made it necessary to employ information from high-energy collider data to constrain the global data fit in the region that was not covered in the CLAS experiment. 
Consequently, large systematic uncertainties were assigned to the results. 

New experimental results on DVCS measurements with polarized electron beams at higher energy have recently been published from  experiments with CLAS12~\cite{CLAS:2022syx} and from Hall A at Jefferson Laboratory~\cite{JeffersonLabHallA:2022pnx}. They extend the kinematic reach both to higher and to lower values in $\xi$, and increase the range covered in $Q^2$. The latter will allow for more sensitive measurements of the $Q^2$ evolution of the DVCS cross section. These new data may also support application of machine learning techniques and artificial neural networks in the higher level data analysis as have been developed by several groups~\cite{Kumericki:2019ddg,Berthou:2015oaw,Grigsby:2020auv}.  

Ongoing experiments and future planned measurements that employ proton and deuterium (neutron) targets, spin-polarized transversely to the beam direction, have strong sensitivity to CFF $\mathcal{E}$. Precise knowledge of the kinematic dependence of $\mathcal{E}(\xi,t)$ is needed to measure the quark angular momentum distribution encoded in the GFF $J_q(t)$ of the proton~\cite{Ji:1996ek}, as defined in Sect.~\ref{sec-II.A}.

The plan to extend the Jefferson Lab's electron accelerator energy reach to 22~GeV would more fully open access to employing $J/\Psi$ production near threshold in a wide $t$ range, and some $\xi$ range to access the gluon part $D_G(t)$ of the proton's $D$-term.  

DVCS data from the COMPASS experiment at CERN with 160~GeV of oppositely polarized $\mu^+$ and $\mu^-$ beams~\cite{COMPASS:2018pup} reach to smaller $\xi$ values and into the sea-quark region. The average of the measured $\mu^+$ and $\mu^-$ cross sections allows for the determination of Im$\cal{H}$. Results from high statistics runs that cover the lower $x_B$ domain are expected in the near future. With these new data, the difference of $\mu^+$ and $\mu^-$ cross sections can also be formed to  obtain the charge asymmetry, which provides direct access to Re$\cal{H}$.

A long term perspective is provided by the planned Electron-Ion Collider projects in the US~\cite{AbdulKhalek:2021gbh,Burkert:2022hjz} and in China~\cite{Anderle:2021wcy}. The US project will extend the kinematic reach in $x_B > 10^{-4}$ and thus will cover with high operational luminosity up to $10^{34}\, \text{cm}^{-2}\,\text{s}^{-1}$ the gluon dominated domain. It features polarized electron and polarized proton beams, the latter longitudinally or transversely polarized, and light ion beams. The EIcC in China focuses on the lower energy domain with $x_B > 10^{-3}$ that connects more closely to the kinematics of the fixed target experiments at Jlab that operate at very high luminosity in the valence quark and the $q\bar{q}$-sea domain. 

Currently available data allowed for a pioneering first step into this emerging new field of the proton internal structure, complementing what has been learned in many detailed experiments over the past 70 years of studies of the proton electromagnetic structure, with the first result on the proton's mechanical structure. 

\begin{figure}[t!]
\includegraphics[width=1.0\columnwidth]{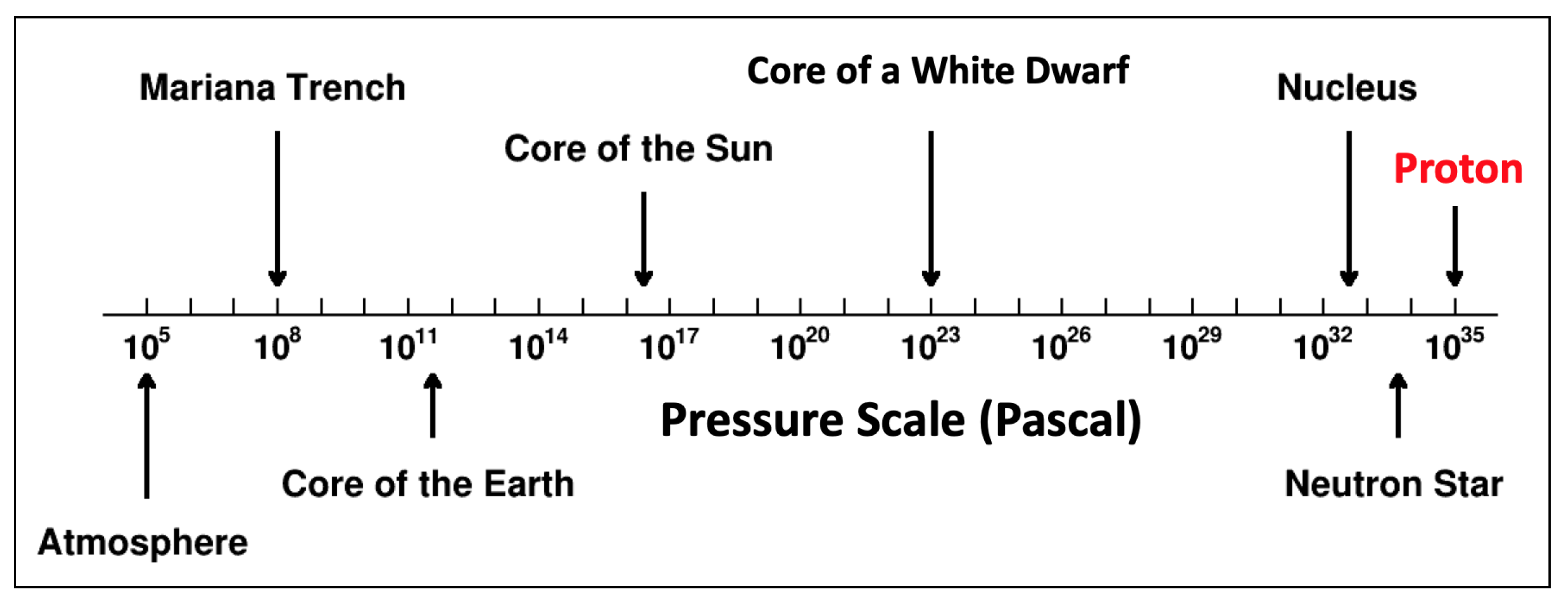}
\caption{\footnotesize Comparison of peak pressures inside various objects on earth, in the solar system, and in the universe.}
\label{PressureScale}
\end{figure} 
 
This new avenue of research has been rapidly developing theoretically, and the first experimental results on the proton firmly established the study of mechanical properties of sub-atomic particle as an exciting new field of fundamental science. Many objects on earth, in the solar system and in the universe are described by their equation of state, where the internal pressure plays an essential role. Some of these objects are listed in Figure~\ref{PressureScale}. The study discussed in this Colloquium adds the smallest object with the highest internal pressure to this list of objects that have been studied so far. The peak pressure inside the proton is approximately $10^{35}$ Pascal. It tops by 30 orders of magnitude the atmospheric pressure on earth. It even exceeds the pressure in the core of the most densely packed known macroscopic objects in the universe, neutron stars, which is given as $1.6\times 10^{34}$ Pascal in \cite{Ozel:2016oaf}. Other subatomic objects such as pions, kaons, hyperons, and light and heavy nuclei may be subject of experimental investigation in the future. The scientific instruments needed to study them efficiently are in preparation. 

The gravitational form factors
provide the key to address fundamental questions 
about the mass, spin, and internal forces 
inside the proton and other hadrons. 
Theoretical, experimental and phenomenological
studies of gravitational form factors provide exciting insights. 
In this emerging new field, there are many
inspiring lessons to learn and there is much to
look forward to.

\begin{acknowledgments}

The authors cannot mention the names of all
colleagues and acknowledge all discussions
during and prior to the preparation of this
Colloquium. 
We wish to name only Maxim Polyakov 
who until his untimely death in August 2021
influenced or initiated many of the reviewed~topics. \\
Special thanks go to Joanna Griffin 
for professional assistance with 
preparing diagrams and figures.
This material is based upon work supported by the U.S. Department of Energy, Office of Science, Office of Nuclear Physics under contract DE-AC05-06OR23177.
P.~Schweitzer was supported by the NSF grant under the Contract No.~2111490. 
P.~Shanahan was supported in part by the U.S.\ Department of Energy, Office of Science, Office of Nuclear Physics, under grant Contract Number DE-SC0011090 and by Early Career Award DE-SC0021006, by a NEC research award, and by the Carl G and Shirley Sontheimer Research Fund. This work was supported in part by the Department of Energy within framework of the QGT Topical Collaboration.
\end{acknowledgments}

\section*{Acronyms}
{A heavy use of acronyms can make a text 
difficult to read for readers not familiar 
with the field, while no use of acronyms can 
make it unreadable for those who are familiar.  
The authors found it indispensable to
introduce a number of acronyms which are
explained at their first occurrence and
are collected here for convenience.

\ \\
\begin{tabular}{ll}
{\it AM}    & angular momentum\\
{\it BSA}   & beam spin asymmetry\\
{\it CFF}   & Compton form factor \\
{\it DIS}   & deep inelastic scattering \\
{\it DVCS}  & deeply virtual Compton scattering \\
{\it EMT}   & energy momentum tensor \\
{\it FF}    & form factor \\
{\it GFF}   & gravitational form factor \\
{\it GPD}   & generalized parton distribution \\
{\it JLab}  & Jefferson Lab \\
{\it PDF}   & parton distribution function \\
{\it QCD}   & quantum chromodynamics \\
{\it QED}   & quantum electrodynamics \\ 
{\it TCS}   & time-like Compton scattering \\
\end{tabular}

}

\bibliography{refs}

%
%

\end{document}